\documentclass[amsmath,amssymb,aps]{revtex4}

\selectfont 
\usepackage{graphicx} 
\usepackage{dcolumn} 
\usepackage{bm} 
\usepackage{color}
\usepackage{multirow}
\usepackage{float} 
\usepackage{verbatim}
\def\e{\mathrm{e}}

\def\be{\begin{equation}}
\def\ee{\end{equation}}
\def\bea{\begin{eqnarray}}
\def\eea{\end{eqnarray}}
\def\L{\left}
\def\R{\right}

\begin{document}

\title{Charged fermion in $(1+2)$-dimensional wormhole with axial magnetic field }

\author{Trithos Rojjanason}

\email{trithot@hotmail.com}

\affiliation{High Energy Physics Theory Group, Department of Physics, Faculty of Science, Chulalongkorn University, Bangkok 10330, Thailand}

\author{Piyabut Burikham}

\email{piyabut@gmail.com}

\affiliation{High Energy Physics Theory Group, Department of Physics, Faculty of Science, Chulalongkorn University, Bangkok 10330, Thailand}

\author{Kulapant Pimsamarn}

\email{fsciklpp@ku.ac.th}

\affiliation{Department of Physics, Faculty of Science, Kasetsart University, Bangkok 10900, Thailand}

\date{\today}

\begin{abstract}

We investigate the effects of magnetic field on a charged fermion in a $(1+2)$-dimensional wormhole.  Applying external magnetic field along the axis direction of the wormhole, the Dirac equation is set up and analytically solved in two scenarios, constant magnetic flux and constant magnetic field through the throat of the wormhole.  For the constant magnetic flux scenario, the system can be solved analytically and exact solutions are found.  For the constant magnetic field scenario, with the short wormhole approximation, the quantized energies and eigenstates are obtained.  The system exhibits both the spin-orbit coupling and the Landau quantization for the stationary states in both scenarios.  The intrinsic curvature of the surface induces the spin-orbit and spin-magnetic Landau couplings that generate imaginary energy.  Imaginary energy can be interpreted as the energy dissipation and instability of the states.  Generically, the states of charged fermion in wormhole are quasinormal modes~(QNMs) that could be unstable for positive imaginary frequencies and decaying for negative imaginary ones.  For the constant flux scenario, the fermions in the wormhole can behave like bosons and have arbitrary statistics depending on the flux.  We also discuss the implications of our results in the graphene wormhole system.

\end{abstract}

\maketitle

\section{Introduction}\label{sec:intro}

A quantum particle can be subject to various kinds of constraints resulting in a number of interesting phenomena.  One of such constraints is the confinement to a surface.  When a quantum particle is confined to a curved surface, its quantum behaviour is nontrivially influenced by the curvatures of the space.  A pioneering work for non-relativistic quantum mechanics in curved two dimensional surface is investigated in Ref.~\cite{daC1,daC2} where the effects of extrinsic and intrinsic curvature are explored.  Generalization to include spin and relativistic effects using Dirac equation reveals a number of interesting consequences of the confinement and curvature of the confining surface.  Notably, electron in graphene~\cite{Novoselov01,Novoselov02} near the Dirac points can be described as fermionic quasiparticle obeying massless Dirac-like equation~\cite{Zhang,Yin,Castro,Rozhkov,Rashba}.

Applying gauge field to the constrained quantum particles can generate curious effects.  Charged particles in a surface attain Landau quantization when a magnetic field is applied in the normal direction.  Even when the particle moves in the region with zero field, it can still experience phase shift when travelled around the non-zero field region, i.e., the Aharonov-Bohm~(AB) effect~\cite{AB}.  The AB effect also occurs when the charged particles are confined to a surface such as the nanotube~\cite{AjAn}.  Quantum Hall effects are notably an example of profound phenomena emerging in the constrained fermionic system with external gauge fields~(see e.g. Ref.~\cite{Tong} and references therein).  Mechanical strain can also mimic effects of the gauge field.  For example, electrons in deformed nanotube and graphene experience deformed potential generated from the strain tensors~\cite{Suzuura:2002ex,Guinea:2009vd,DeJuan:2013pha}.

There is a number of investigation of fermions confined to a curved surface~\cite{BuJe,Hansson,Entin,Wang:2014,Wang:2017,Liang} as well as the implications to carbon nanotubes properties~\cite{AjAn,Varsano}, and applications in curved graphene~\cite{Cariglia,Thitapura,Biswas,Lherbier,Villarreal,Jakubsky}.  Graphene is an ideal place to study behaviour of confined charged fermions such as electrons in a two-dimensional surface since its thickness is only roughly one-carbon-atom diameter.  A sheet of graphene can be curved, rolled, stretched, twisted and deformed or even punctured holes into.  The holes can be connected to a nanotube and become a wormhole bridging two graphene sheets.  Multiple graphene sheets can be connected with one another by multiple wormholes forming a network of entangled electronic structure.  Wormholes can even be built into a cage structure of schwarzite with many promising properties~\cite{schw1}.  

There have been many studies concerning the behavior of electron on curved graphene surfaces.  Gonzalez et al.\cite{Gonzalez} consider a wormhole attached to two graphene sheets via 12 heptagonal defects, the defects act like effective non-Abelian gauge flux that swaps two Dirac points on the graphene lattice.  Garcia et al.\cite{Garcia} investigate the charged fermion in two-dimensional spherical space in a rotating frame, study the change in the spectrum of the $C_{60}$ molecule when it is crossed by a magnetic flux tube in the $z$-direction, and the appearance of an analogue of the Aharonov-Carmi phase in the system~\cite{Shen}.  Cariglia et al.\cite{Cariglia} consider Dirac fermions on an essentially smooth simplified spacetime, namely a Bronnikov-Ellis wormhole.  In Ref.~\cite{IoLa}, the surface of the graphene wormhole is realized by a two-dimensional axially-symmetric curved space of a constant Gaussian curvature.  The effective action of the fermion in the graphene wormhole is then derived in the (1+2)-dimensional spacetime. The similarities~(in the long wormhole limit) and differences of Hilbert and event horizon are discussed.    In Ref.~\cite{Cvetic:2012vg}, a charged fermion in curved surface subject to external electric field is analyzed in the stationary optical metric conformal to the BTZ black hole.  Rojjanason and Boonchui \cite{Rojjanason} showed that the exact solution of the Dirac fermion on the graphene wormhole can be expressed in terms of Jacobi polynomials and the spin-orbit coupling is generated by the curvature of the wormhole. It should be emphasized that the ``wormhole'' in these condensed matter system is different from astrophysical wormhole in two crucial aspects.  Firstly, the condensed matter wormhole is basically spatially curved 2-dimensional surface, so it is $(1+2)$-dimensional.  Secondly, there is no time dilatation like in the astrophysical wormhole, it is assumed that time is not affected by the curved surface.  

In this work, we study physical properties of a charged fermion confined on the surface of wormhole in the presence of the external magnetic field along the axis direction of the wormhole. In Section \ref{sec:Line element}, basic geometric and gauge setup are established.  In Section \ref{sec:Dirac eq.}, the Dirac equation in curved spacetime is used to analyze the (1+2)-dimensional stationary state of the charged fermion in the wormhole.  To solve for the energy and wave function, two scenarios of constant flux and constant field are considered.  Analysis in special cylindrical wormhole case is compared to the general case to identify the crucial role of surface curvature.  A simple interpretation of the results is given in terms of the angle between the spin and orbital angular momentum of the surface-confined fermion.  The special cases of Beltrami and elliptic wormhole are considered in Section \ref{BelEll}.  Implications to the graphene system are discussed in Section~\ref{graSec}.  We summarize and discuss our results in Section \ref{sec:con}.

\section{\label{sec:Line element}Geometric and gauge setup of the wormhole}

\begin{figure}[h!]
  \centering
  \includegraphics[width=0.85\textwidth]{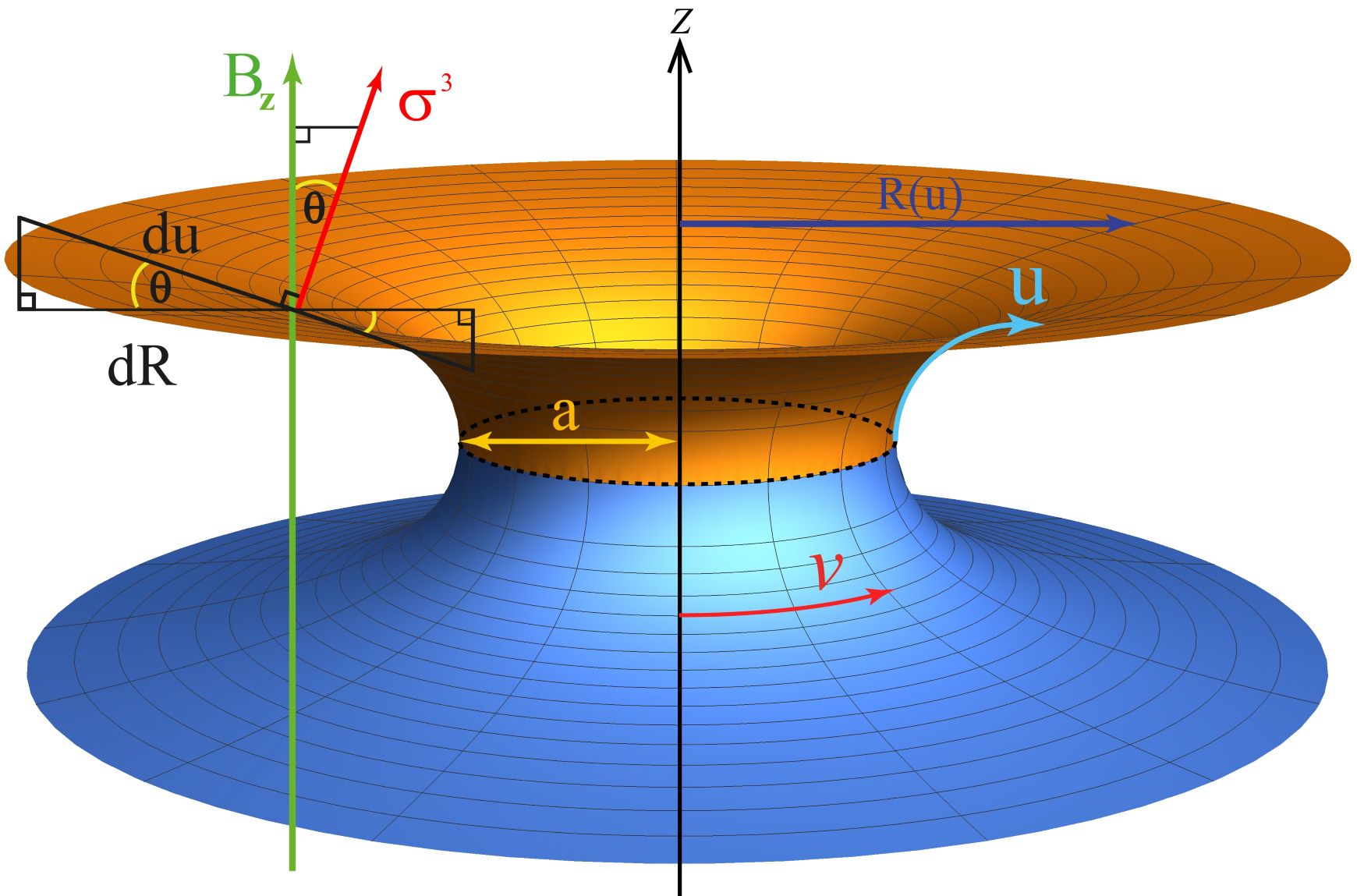}
  \caption{Geometric structure of a wormhole surface where $a$ is a radius of the wormhole at the mid-point, $u=0$, between the two ends.  And $r$ is the radius of curvature of the wormhole surface along $u$ direction. }  \label{fig1}
\end{figure}

The wormhole is defined geometrically as in Figure~\ref{fig1} with points on the surface parameterized by
\begin{equation} \label{eq:1}
\vec{\textbf{r}}(u,v)=x(u,v)\hat{i} + y(u,v)\hat{j} + z\hat{k}
\end{equation}
where
\begin{equation}\nonumber
x(u,v)=R(u)\cos(v), ~y(u,v)= R(u)\sin(v), \\
z(u)= \int du \,\,\sqrt{1-\left(R'(u)\right)^{2}}.
\end{equation}
The shape of the wormhole is generically described by the radius function $R(u)$.  The constraint on $z$ follows from the relation $ds^2 = dx^{2}+dy^{2}+dz^{2}=du^{2}+R(u)^{2}dv^{2}$.  It gives the Hilbert horizon at $1=R'(u)$~\cite{IoLa}.

Embedding $(1+2)$-D wormhole into higher dimensional $(1+3)$-D space generates effective gravity or effective curvature to the reduced ``spacetime''.  Any particle or quasiparticle living on the reduced spacetime will experience the spacetime curvature.
We define $dx^{\mu'}=\{cdt,dx,dy,dz\}$ as the (1+3)-D Minkowski spacetime coordinates, and $dx^{\mu}=\{cdt,du,dv\}$ as the (1+2)-D wormhole coordinates.  The transformation matrix between the two coordinates is then 
\begin{equation}\nonumber
\frac{\partial_{}x^{\mu'}}{\partial_{}x^{\nu}}=\begin{pmatrix}
1&0&0&0\\
0&R'(u)\cos(v)&R'(u)\sin(v)&\sqrt{1-\left(R'(u)\right)^{2}}\\
0&-R(u)\sin(v)&R(u)\cos(v)&0\\
  \end{pmatrix}.
\end{equation}

Since the line element of the wormhole space is in the following form
\begin{equation}\label{eq:7}
ds^{2}=-c^{2}dt^{2}+du^{2}+R^{2}(u)dv^{2}= g_{\mu \upsilon} dx^{\mu} dx^{\upsilon},
\end{equation}\\
where the $(1+2)$-D metric is $g_{\mu \upsilon}$.  The dreibein $e^{A}_{\mu}$ is then defined as
\begin{equation}\label{eq:8}
e^{A}_{\mu}=\begin{pmatrix}
1&0&0\\
0&1&0\\
0&0&R(u)\\
  \end{pmatrix},
\end{equation}\\
where $g_{\mu \nu}\equiv e^{A}_{\mu}e^{B}_{\nu}\eta_{AB}$, and $\eta_{AB}=diag(-1,1,1)$ in (1+2)-dimensions and $ A,B\in \{0,1,2\}$.
We consider an electron (or charged fermion) in Minkowski space subject to the wormhole embedding constraints.  The fermion will experience the effective curvature that can be addressed by considering the Dirac equation in the locally flat $(1+2)$-D spacetime
\begin{equation}\nonumber
\Big[\gamma^{A}p_{A}-Mc\Big]\Psi = 0.
\end{equation}
The canonical momentum of fermion in the presence of an electromagnetic potential is
\begin{equation}\nonumber
p_{A}=e^{\mu}_{A}(-\hbar\nabla_{\mu}+i\frac{e}{c}A_{\mu}).
\end{equation}
therefore the coordinate-space Dirac equation can be written as
\begin{equation}\label{eq:9}
\Big[\gamma^{A}e^{\mu}_{A}(-\hbar\nabla_{\mu}+i\frac{e}{c}A_{\mu})-Mc\Big]\Psi = 0.
\end{equation}
$\Psi=\Psi(u,v,t)$ represents the Dirac spinor field and $M$ represents the rest mass of the particle, $c$ is the speed of light in the curved spacetimes, $e$ is electric charge, and $A_{\mu}$ is the electromagnetic four-potential.  The $\gamma^{A}$ are the Dirac matrices given by
\begin{equation}\nonumber
\gamma^{0}=\left(
 \begin{array}{cc}
 i & 0 \\
  0 & -i \\
  \end{array}
 \right)
,\quad
\gamma^{a}=\left(
  \begin{array}{cc}
  0 & i\sigma^{a} \\
  -i\sigma^{a} & 0 \\
  \end{array}
 \right),
\end{equation}
where $\sigma^{a}$ are the Pauli matrices defined by
\begin{equation}\nonumber
 \sigma^{1}=\left(
 \begin{array}{cc}
 0 & 1 \\
  1 & 0 \\
  \end{array}
 \right)
,\quad
\sigma^{2}=\left(
  \begin{array}{cc}
  0 & -i \\
  i & 0 \\
  \end{array}
 \right),\quad
 \sigma^{3}=\left(
  \begin{array}{cc}
  1 & 0 \\
  0 & -1 \\
  \end{array}
\right).
\end{equation}
They obey the Clifford algebra
\begin{equation}\label{eq:10}
\{\gamma^{A}, \gamma^{B}\}=2\eta^{AB}.
\end{equation}
The Pauli matrices satisfy a useful identity that we will use later
\begin{equation}\label{eq:11}
\sigma^{a}\sigma^{b}=\delta^{ab}+i\epsilon^{abc}\sigma^{c},
\end{equation}
where $\epsilon^{abc}$ is Levi-Civita symbol.

The covariant derivative of the spinor interaction with gauge field in the curved space is given by
\begin{equation} \label{eq:12}
\nabla_{\mu}\equiv \partial_{\mu}-\Gamma_{\mu},
\end{equation}
where the spin connection $\Gamma_{\mu}$ \cite{Yepez} is
\begin{equation}\label{eq:11}
\Gamma_{\mu}=-\frac{1}{4} \gamma^{A}\gamma^{B} e^{\nu}_{A} \Big(\partial_{\mu}(g_{\nu \beta}e^{\beta}_{B})-e^{\beta}_{B}\Gamma_{\beta\mu\nu} \Big),
\end{equation}
where $\beta,\mu,\nu \in \{t,u,v\}$ and the Christoffel symbols $\Gamma_{kji}$ are defined by
\begin{equation}\nonumber
\Gamma_{\beta \mu \nu}=\frac{1}{2}(\partial_{\mu}g_{\beta \nu}+\partial_{\nu}g_{\beta \mu}-\partial_{\beta}g_{\mu \nu}).
\end{equation}
Then
\begin{equation}\label{eq:12}
-\Gamma_{uvv} = \Gamma_{vuv} = \Gamma_{vvu}=\frac{1}{2}\partial_{u}R^{2},
\end{equation}

and zero otherwise. The spin connections are then
\begin{equation}\label{eq:13}
\begin{split}
\Gamma_{t}
         &=-\frac{1}{4} \gamma^{A}\gamma^{B} e^{\nu}_{A} \Big(\partial_{t}(g_{\nu \beta}e^{\beta}_{B}) - e^{\beta}_{B}\Gamma_{\beta\nu t} \Big)=0   \\
\Gamma_{u}
         &=-\frac{1}{4} \gamma^{A}\gamma^{B} e^{\nu}_{A} \Big(\partial_{u}(g_{\nu \beta}e^{\beta}_{B}) - e^{\beta}_{B}\Gamma_{\beta\nu u} \Big)   \\
         &=-\frac{1}{4} \gamma^{2}\gamma^{2} e^{v}_{2}\partial_{u}( g_{vv}e^{v}_{2})+\frac{1}{4} \gamma^{2}\gamma^{2} e^{v}_{2} e^{v}_{2}\Gamma_{vvu}=0   \\
\Gamma_{v}
         &=-\frac{1}{4} \gamma^{A}\gamma^{B} e^{\nu}_{A} \Big(\partial_{v}(g_{\nu \beta}e^{\beta}_{B}) - e^{\beta}_{B}\Gamma_{\beta\nu v} \Big)   \\
         &=\frac{1}{4} \gamma^{1}\gamma^{2} e^{u}_{1} e^{v}_{2}\Gamma_{vuv} + \frac{1}{4} \gamma^{2}\gamma^{1}e^{v}_{2}e^{u}_{1}\Gamma_{uvv}=\frac{1}{2}\gamma^{1}\gamma^{2}R'.          
\end{split}\end{equation}
In this work, we will apply an external magnetic field such that the $z$-component $B_{z}=B(z)$ is uniform with respect to the plane $(x,y)$ in two different ways: a.) the magnetic {\it flux} through the circular area enclosed by the wormhole at a fixed $z$ is constant, namely $B_{z}\sim 1/R^{2}$ and b.) the magnetic {\it field} is uniform and constant. Due to the axial symmetry, the electromagnetic four-potential can be expressed in the axial gauge as
\begin{equation}\nonumber
A_{\mu'}(t,x,y,z)=(0,-\frac{1}{2}By,\frac{1}{2}Bx,0),
\end{equation}
and in the wormhole coordinates as
\begin{equation}\nonumber
A_{\mu}(t,u,v)=\frac{\partial_{}x^{\nu'}}{\partial_{}x^{\mu}}A_{\nu'}(t,x,y,z).
\end{equation}
Specifically by components, they are
\begin{equation}\label{eq:16}
A_{t}=0, A_{u}=\frac{\partial_{}x}{\partial_{}u}A_{x}+\frac{\partial_{}y}{\partial_{}u}A_{y}=0, A_{v}=\frac{\partial_{}x}{\partial_{}v}A_{x}+\frac{\partial_{}y}{\partial_{}v}A_{y}=\frac{1}{2}BR^{2}.   
\end{equation}
The magnetic field is then given by
\be
\vec{B} = (-\frac{x}{2}\partial_{z} B, -\frac{y}{2}\partial_{z} B, B),
\ee
having all $x, y, z$ components for the constant-magnetic-flux case and having only $z$ component for the constant-magnetic-field case.

\section{\label{sec:Dirac eq.}The Dirac equation in magnetized wormhole}

Utilizing the results from above equations, the Dirac equation Eq.(\ref{eq:9}) can be written in the form
\begin{equation}\nonumber
\Big[\gamma^{0}e^{t}_{0}(-\hbar\nabla_{t}+i\frac{e}{c}A_{t})+\gamma^{1}e^{u}_{1}(-\hbar\nabla_{u}+i\frac{e}{c}A_{u})+\gamma^{2}e^{v}_{2}(-\hbar\nabla_{v}+i\frac{e}{c}A_{v})-Mc\Big]\Psi=0,
\end{equation}
leading to
\be
\Big[\gamma^{0}\partial_{ct}+\gamma^{1}\partial_{u}+\gamma^{2}\Big(\frac{1}{R}\Big)(\partial_{v}-\frac{1}{2}\gamma^{1}\gamma^{2}R'-\frac{ie}{2\hbar c}BR^{2})+\frac{Mc}{\hbar}\Big]\Psi=0.
\ee
By using relationships from Eq.(\ref{eq:10})
\begin{equation}\label{eq:17}
\Big[\gamma^{0}\partial_{ct}+\gamma^{1}\Big(\partial_{u}+\frac{R'}{2R}\Big)+\gamma^{2}\Big(\frac{1}{R}\partial_{v}-\frac{ie}{2\hbar c}BR\Big)+\frac{Mc}{\hbar}\Big]\Psi=0,
\ee
then
\be
\left(\begin{array}{cc}
 (i\partial_{ct}+\frac{Mc}{\hbar}) & i\textbf{D} \\
  -i\textbf{D} & (-i\partial_{ct}+\frac{Mc}{\hbar}) \\
  \end{array}\right)\Psi=0,   \label{eom}
\end{equation} 
where $\textbf{D}$ is a differential operator 
\begin{equation}\label{eq:21}
\textbf{D}\equiv \sigma^{1}\Big(\partial_{u}+ \frac{R'}{2R}\Big)+ \sigma^{2}\Big(\frac{1}{R}\partial_{v}-\frac{ie}{2\hbar c}BR\Big).
\end{equation}
We can define the pseudo vector potential as
\begin{equation}
\textbf{D}=\sigma^{1}\Big(\partial_{u}-i\frac{e}{\hbar c}A_{\tilde{u}}\Big)+ \sigma^{2}\frac{1}{R}\Big(\partial_{v}-i\frac{e}{\hbar c}A_{v}\Big)\quad    \\
;\quad A_{\tilde{u}}\equiv i\frac{\hbar c}{e}\frac{R'}{2R},\quad A_{u}=0,\quad A_{v}=\frac{1}{2}BR^{2}.  \label{coneq}
\ee
The effective gauge potential in the $u$ direction, $A_{\tilde{u}}$, is generated by the curvature along the $v$ direction, $\Gamma_{v}$.  In this sense, wormhole gravity connection manifests itself in the form of (imaginary) gauge connection in the perpendicular direction on the surface.  The second term in Eq.(\ref{eq:21}) is similar to a spin-orbit-curvature coupling potential. A similar setup has been used to study nanotubes under a sinusoidal potential \cite{Gonzalez}.  Here we consider the dispersion relation for the two-dimensional fermions described by the Dirac equation in the presence of the effective potential arising from the wormhole geometrical structure and the gauge field.

Consider a stationary state of the Dirac spinor $\Psi(t,u,v)$ in the form
\begin{equation}\label{eq:18}
\Psi(t,u,v)=e^{-\frac{i}{\hbar}Et}\left(
\begin{array}{cc}
  \psi_{I}(u,v) \\
  \psi_{II}(u,v) \\
  \end{array}
\right ),
\end{equation}

where $\psi_{I(II)}(u,v)$ are two-component spinors.  Eq.(\ref{eom}) can be rewritten in the form of coupled equations for the 2-spinors
\begin{equation}\label{eq:19}
\left(\frac{E}{\hbar c}+\frac{Mc}{\hbar}\right)\psi_{I}(u,v)+i\textbf{D}\psi_{II}(u,v) =0,
\end{equation}
\begin{equation}\label{eq:20}
\left(-\frac{E}{\hbar c}+\frac{Mc}{\hbar}\right)\psi_{II}(u,v)-i\textbf{D}\psi_{I}(u,v) =0.
\end{equation}

In the presence of external magnetic field $\mathbf{B}=\bigtriangledown\times\mathbf{A}$ along the $z$ direction, the charged fermion moving in $v$ direction is expected to form a stationary state with quantized angular momentum and energy, i.e. the Landau levels in the curved space with hole.  To show this, we need to solve for the stationary states of the system.  Consider $-i\textbf{D}\times$Eq.(\ref{eq:19})-$(\frac{E}{\hbar c}+\frac{Mc}{\hbar})\times$Eq.(\ref{eq:20}) to obtain
\begin{equation}\label{eq:22}
\begin{split}
\Big[(-i\textbf{D})(\frac{E}{\hbar c}+\frac{Mc}{\hbar})-(\frac{E}{\hbar c}+\frac{Mc}{\hbar})(-i\textbf{D})\Big]\psi_{I}(u,v)+\Big[\textbf{D}^{2}-(\frac{E}{\hbar c}+\frac{Mc}{\hbar})(-\frac{E}{\hbar c}+\frac{Mc}{\hbar})\Big]\psi_{II}(u,v)
              &=0,   \\
\Big[\textbf{D}^{2}+\frac{E^2-M^2c^4}{\hbar^2 c^2}\Big]\psi_{II}(u,v)
              &=0.   \\
\end{split}
\end{equation}
Now we will solve the equation of motion in two cases.  We will demand the wave function to be regular in the wormhole and will not specify the boundary condition at both ends~(i.e. Hilbert horizons) of the wormhole.  This corresponds to the physical situations where the wormhole is connected between two upper and lower surfaces whereby the wave function needs not to be zero.   

\subsection{Constant magnetic flux solution}

์In type II superconductors, the magnetic flux can be trapped in a magnetic vortex~(i.e. Abrikosov vortex) in quantized units of the magnetic flux quantum $\phi_{0}\equiv hc/e$.  The flux will be kept constant along the vortex which can be thought of as a wormhole if the charged carriers are confined to the boundary of the vortex by other constraint.  For example, the confining surface can be created as the wormhole connecting two graphene sheets.  In such situation, our analysis in this section would be applicable.  The generic results demonstrate that the statistics of charged fermion in the wormhole is determined solely by the magnetic flux and not the shape or curvature of the wormhole via the effective orbital angular momentum.  The fermion can behave like boson when the flux is half-integer and have arbitrary statistics for arbitrary fluxes.       

The presence of axial magnetic field along the wormhole gives rise to a number of interesting effects notably the Landau quantization of energy due to the interaction between charge of the fermion and the external magnetic field.  On the other hand, the wormhole {\it intrinsic} curvature~(``gravity'') is responsible for the spin-orbit coupling of the fermion as we will see subsequently in this Section and later on in Section \ref{conR}~(where the spin-orbit disappears with the {\it intrinsic} curvature).  It costs energy to tilt the spin along the wormhole space even in the absence of the magnetic field. 

We will start with the situation where the magnetic {\it flux} is constant along the throat of the wormhole. 

The operator $\textbf{D}^{2}$ in the equation of motion now takes the form
\begin{equation}\label{Dsq}
\begin{split}
\textbf{D}^2
          &=\sigma^{1}\sigma^{1}\Big(\partial_{u}+ \frac{R'}{2R}\Big)^2+ \sigma^{2}\sigma^{2}\Big(\frac{1}{R}\partial_{v}-\frac{1}{R}i\frac{\phi}{\phi_0}\Big)^2   \\
          &\quad+\sigma^{1}\sigma^{2}\Big(\partial_{u}+ \frac{R'}{2R}\Big)\Big(\frac{1}{R}\partial_{v}-\frac{1}{R}i\frac{\phi}{\phi_0}\Big)+\sigma^{2}\sigma^{1}(\frac{1}{R}\partial_{v}-\frac{1}{R}i\frac{\phi}{\phi_0}\Big)\Big(\partial_{u}+ \frac{R'}{2R}\Big)    , \\
          &=\partial_{u}^2
          +\frac{R'}{R}\partial_{u}
          -\Big(\frac{R'}{2R}\Big)^2
          +\frac{R''}{2R}
          +\frac{1}{R^2}\Big(\partial_{v}-i\frac{\phi}{\phi_0}\Big)^2
          -i\sigma^{3}\frac{R'}{R^2}\Big(\partial_{v}-i\frac{\phi}{\phi_0}\Big)
\end{split}
\end{equation} 
where $\phi =\int \vec{B}\cdot d\vec{a}=\pi R^{2}B$ and $\phi_{0}$ the magnetic flux quantum.  We have used the identity (\ref{eq:11}), $\sigma^{1}\sigma^{2}=i\sigma^{3}$.  For zero magnetic field, the operator $\textbf{D}^2$ satisfies the Lichnerowicz-Weitzenb$\ddot{\rm o}$ck formula
\bea
\textbf{D}^{2} &=& \nabla^{2} + \frac{1}{4}\mathcal{R},  \\
                    &=& \frac{1}{\sqrt{-g}}D_{\mu}\left[ \sqrt{-g}~g^{\mu\nu}D_{\nu} \right]+ \frac{1}{4}\mathcal{R}
\eea
where $D_{\mu}=\partial_{\mu}-\Gamma_{\mu}$ is the covariant derivative including the spin connection and $\mathcal{R}$ is the Ricci scalar.  In our case, $\mathcal{R}=2R''/R.$  

For stationary states, the wave function needs to be single-valued at every point in spacetime, $\psi_{II}$ must be a periodic function in $v$,
\begin{equation}\label{varsep}
\Phi_{II}(u,v)=e^{i m v}\left(
\begin{array}{cc}
\varphi^+_{II}(u)\\
 \varphi^-_{II}(u) \\
\end{array}
\right). 
\end{equation}
where the orbital angular momentum quantum number $m=0,\pm 1,\pm 2,...$

 Substitute Eq.(\ref{Dsq})-Eq.(\ref{varsep}) into Eq.(\ref{eq:22}) to obtain 
\begin{equation}\label{eq:25}
\begin{split}
0
          &=\Big[\textbf{D}^2+k^2\Big] \varphi^{\sigma}_{II}(u),    \\
0
          &=\Big[\partial_{u}^2
          +\frac{R'}{R}\partial_{u}
          +\frac{R''}{2R}
          +\frac{m'\sigma R'-m'^2-\Big(\frac{R'}{2}\Big)^2}{R^2}+k^2\Big] \varphi^{\sigma}_{II}(u)
\end{split}
\end{equation} 
where $m'=m-\displaystyle{\frac{\phi}{\phi_{0}}}$, the new orbital angular momentum in the presence of magnetic flux~\cite{Wilczek}.  Notably by adjusting the flux $\phi/\phi_{0}$ to be half an odd integer, the charged fermions such as electrons in the wormhole can behave like bosons, e.g. they can condensate and flow like superfluid along the wormhole.  We have used the momentum parameter $k^{2} \equiv (E^{2}-M^{2}c^{4})/\hbar^{2}c^{2}$ and $\sigma$ is a spin-state index corresponding to spin up ($\sigma$=+1) or down ($\sigma=-1$) of the fermion for each eigenvalue of $\sigma^3$.  The $\sigma^{3}$-spin component is pointing along the direction of the normal vector of the wormhole surface since the dreibein $e^{A}_{\mu}$ is defined on the tangent space of the wormhole.  Also because $R'(u)=\cos\theta$ where $\theta$ is the angle between the $\sigma^{3}$-spin component and the $z$ axis, the spin-orbit coupling term can thus be rewritten as $\sim \sigma m R'=\sigma m\cos\theta=\overrightarrow{\sigma^{3}}\cdot \hat{z}m$.  The spin-orbit coupling vanishes when $R'=0=\cos\theta$ or $\theta =\pi/2$, i.e. when the normal vector of the surface is perpendicular to $\hat{z}$.

To be specific, we will choose a deformed hyperbolic wormhole described by $R(u)=a\cosh_q(u/r)$ where $a$ is the radius of wormhole at the mid-point between the two sheets and $r$ is the radius of curvature of the wormhole connecting the sheets~\cite{IoLa}. They are based on a $q$-deformation of the usual hyperbolic functions which are defined by \cite{Arai,Egrifes}
\begin{equation} \label{eq:2}
\cosh_q(x)\equiv\frac{e^{x}+qe^{-x}}{2}, \quad \sinh_q(x)\equiv\frac{e^{x}-qe^{-x}}{2}, \quad \tanh_q(x)=\frac{\sinh_q(x)}{\cosh_q(x)}.
\end{equation}\\
Definitions same to else hyperbolic functions but note that almost all relations known from the usual hyperbolic functions have been modified, for example

\begin{equation} \label{eq:3}
\cosh_q^2(x)-\sinh^2_q(x)=q, \quad \frac{d}{dx}\sinh_q(x)=\cosh_q(x), \quad \frac{d}{dx}\tanh_q(x)=\frac{q}{\cosh^{2}_q(x)}.
\end{equation}
\\
They reduce to hyperbolic functions when $q=1$.  In this specific choice of $R$, the Gaussian curvature 
\begin{equation}
\kappa = -\frac{R''}{R}= -\frac{1}{r^{2}},
\end{equation}
is a negative constant~\cite{IoLa}.

With this choice of $R(u)$, we perform the transformation $X=\sinh_q(u/r)$ to obtain 
\bea\label{eq:27}
0  &=&\Big(q+X^2\Big)\varphi''(X) +2X\varphi'(X)
         +k^2 r^2\varphi(X)+\Big[\frac{\frac{q}{4}+\frac{r}{a}\sigma m' X-\Big(\frac{r}{a}m'\Big)^2 }{\Big(q+X^2\Big)}
         +\frac{1}{4} \Big]\varphi(X).
\eea   
Define weighting function solution $\varphi(X)=(\sqrt{q}+iX)^\alpha (\sqrt{q}-iX)^\beta\Phi(X)$, the equation of motion can be rewritten as
\begin{equation}
\begin{split}
0
         &=(q+X^2)\Phi''(X) +2\Big[(\alpha+\beta+1)X+i(\alpha-\beta)\sqrt{q}\Big]\Phi'(X)
         +k^2 r^2\Phi(X)                                                              \\
         &\quad +\Big[(\alpha+\beta)(\alpha+\beta+1)+\frac{1}{4}
         \Big]\Phi(X),                                                                \\
0
         &=(q+X^2)\Phi''(X) +\Big[\textup{A}+\textup{B}X\Big]\Phi'(X)
         +\Big[\textup{C}+k^2 r^2 \Big]\Phi(X),\label{eq:32}
\end{split}
\end{equation}
where we assume
\begin{equation}\label{eq:30}
2i\Big(\alpha^2-\beta^2\Big)\sqrt{q}+\frac{r}{a}\sigma m'=0, 
\qquad -2\Big(\alpha^2+\beta^2\Big)q+\frac{q}{4}-\Big(\frac{r}{a}m'\Big)^2=0,  \nonumber
\ee
leading to
\be
\alpha^2=\Big(\frac{1}{4}+\frac{i}{\sqrt{q}}\frac{\sigma m'r}{2a}\Big)^2, 
\qquad \beta^2=\Big(\frac{1}{4}-\frac{i}{\sqrt{q}}\frac{\sigma m'r}{2a}\Big)^2.
\end{equation}
The coefficients are defined by
\begin{equation}\label{eq:31}
\textup{A}
         =2i(\alpha-\beta)\sqrt{q}, \quad\textup{B}=2(\alpha+\beta+1), \quad \textup{C}
         =(\alpha+\beta)(\alpha+\beta+1)+\frac{1}{4}.
\end{equation}
Depending on the sign choices of $\alpha, \beta$, the resulting equation of motion and the corresponding energy levels will be dependent or independent of the spin-orbit coupling term $\sim \sigma mr/a\sqrt{q}$.

Define $X=-i\sqrt{q}Y$, the equation then takes the form
\begin{equation}\label{eq:34}
0=(1-Y^2)\Phi''(Y)+2\Big[(\alpha-\beta)-(\alpha+\beta+1)Y\Big]\Phi'(Y)-\Big[(\alpha+\beta)(\alpha+\beta+1)+k^2 r^2+\frac{1}{4}\Big]\Phi(Y).
\end{equation}
Eq.(\ref{eq:34}) is the Jacobi Differential Equation
\begin{equation}\label{eq:35}
\begin{split}
0
        &=(1-Y^2)\Phi''(Y) +\Big[\beta_0-\alpha_0-(\beta_0+\alpha_0+2)Y\Big]\Phi'(Y)
        +n\Big[n+\beta_0+\alpha_0+1\Big]\Phi(Y)                           \\
\end{split}
\end{equation}
for integer $n$ and
\begin{equation}\label{eq:36}
\beta_0 =2\alpha,\quad \alpha_0=2\beta,\quad k^{2}_{n} r^2 =\frac{E^{2}_n-M^{2}c^{4}}{\hbar^{2}c^{2}}r^2=-\Big(n+\frac{1}{2}+\alpha+\beta \Big)^{2}.
\end{equation}
For the choice
\be
\alpha = \beta^*=\frac{1}{4}+\frac{i}{\sqrt{q}}\frac{\sigma m'r}{2a},
\ee
the energy levels become
\be
E_{n}=\pm \sqrt{M^{2}c^{4}-\left(\frac{\hbar c}{r}\right)^{2}\Big( n+1 \Big)^{2}}, \label{E01}
\ee
independent of the magnetic flux and spin-orbit term.  On the other hand, for another sign choice 
\be
\alpha = -\beta^*=\frac{1}{4}+\frac{i}{\sqrt{q}}\frac{\sigma m'r}{2a},
\ee
the energy levels depend on the spin-orbit term
\be
E_{n}=\pm \sqrt{M^{2}c^{4}-\left(\frac{\hbar c}{r}\right)^{2}\Big( n+\frac{1}{2}+\frac{i}{\sqrt{q}}\frac{\sigma r}{a}(m-\frac{\phi}{\phi_{0}}) \Big)^{2}}, \label{E1}
\ee
a complex quantity which can be interpreted as the quasi-normal modes~(QNMs).  For other combinations with $(\alpha,\beta) \leftrightarrow -(\alpha,\beta)$, the energy becomes
\be
E_{n}=\pm \sqrt{M^{2}c^{4}-\left(\frac{\hbar c}{r}\right)^{2}n^{2}}, \label{E02}
\ee
and 
\be
E_{n}=\pm \sqrt{M^{2}c^{4}-\left(\frac{\hbar c}{r}\right)^{2}\Big( n+\frac{1}{2}-\frac{i}{\sqrt{q}}\frac{\sigma r}{a}(m-\frac{\phi}{\phi_{0}}) \Big)^{2}}  \label{E2}
\ee
respectively.  All of these solutions are part of the Hilbert space of the system with their own corresponding wave functions.  The energy from Eq.~(\ref{E1}) and (\ref{E2}) contains the interaction between the spin-orbit coupling $\sim \sigma m$~(independent of the magnetic field), and the Landau coupling between the magnetic field and the spin~(orbital) angular momentum $\sim \sigma B~(mB)$. Notably, it also contains the term proportional to $\sigma m \phi, m^{2}, \phi^{2}$ which could be interpreted as the spin-orbit-flux, orbit-orbit and the flux-flux couplings respectively.  The spin-orbit and $m^{2}$ terms are purely gravitational and kinematical since they are independent of the magnetic field.  The orbit-orbit $m^{2}$ term is actually the kinetic energy from the angular momentum as we will see in Section \ref{conR} where the curvature vanishes.      

Another interesting aspect of the energies given by Eq.~(\ref{E1}) and (\ref{E2}) is the complexity, i.e., they are quasi-normal modes~(QNMs).  Regardless of the magnetic field, the imaginary parts in the energy expression have the gravitational origin.  They are originated from the curvature of the wormhole and they will vanish when the curvature is zero as we can see again in Section \ref{conR}.  For the QNMs with negative imaginary parts, the curvature effects leak the energy of the fermion away from the wormhole as long as the angle $\theta =\arccos R'$ between the $\sigma^{3}$-spin component and orbital angular momentum is not $\pi/2$.  For the choice (\ref{E1})~(and (\ref{E2})), the positive-energy solution is unstable with positive imaginary part for $\sigma m' <0$~(and $\sigma m'>0$) respectively.  For these states, the fermion will either slowly decay away or be spun off the wormhole due to the curvature effect.  A special case occurs when $m=\phi/\phi_{0}$ where the imaginary spin-orbit coupling term vanishes.    

It is challenging to give physical interpretation to the states from the choice in (\ref{E01}) and (\ref{E02}).  They have negative momentum square $p^{2}$ along $u$ direction, they do not feel the magnetic field and do not have the angular momenta.  It is most natural to identify them with diffusive modes~(due to imaginary momentum along the wormhole direction $u$) with $m=0$.  However, the energy of these modes can be either real or purely imaginary depending on the quantum number $n$.       

The wave-function solutions to Eq.(\ref{eq:35}) are the Jacobi polynomials
\begin{equation}\label{eq:37}
\begin{split}
\Phi_{n}(Y)
          &=P^{(\alpha_{0},\beta_{0})}_{n}(Y)=\frac{(-1)^{n}}{2^{n}n!}(1-Y)^{-\alpha_{0}}(1+Y)^{-\beta_{0}}\frac{d^{n}}{dY^{n}}\Big[(1-Y)^{\alpha_{0}}(1+Y)^{\beta_{0}}(1-Y^2)^{n}\Big]                       \\
 \quad \quad \quad 
          &=\sum_{j=0}^{n}
          	\left(\begin{array}{cc}
			n+\alpha_{0}\\
			n-j\\
			\end{array}\right)
            \left(\begin{array}{cc}
			n+\beta_{0}\\
			j\\
			\end{array}\right)
            \Big(\frac{Y-1}{2}\Big)^{j}\Big(\frac{Y+1}{2}\Big)^{n-j},
\end{split}
\end{equation}
for integer $n$ and
\begin{equation}\nonumber
\left(\begin{array}{cc}
z\\
n\\
\end{array}\right)
=\Bigg{\{}\begin{array}{cc}\begin{split}
\frac{\Gamma(z+1)}{\Gamma(n+1)\Gamma(z-n+1)} \quad  &\text{for} \quad n>0,\\
0 \qquad \quad \qquad \quad &\text{for} \quad n<0.\\
\end{split}\end{array}
\end{equation}
Finally, the solutions of Eq.(\ref{varsep}) is 
\begin{equation}
\Phi_{I/II}(u,v)=e^{-iE_{n}t/\hbar }e^{imv}\Big(\sqrt{q}+iX\Big)^\alpha \Big(\sqrt{q}-iX\Big)^\beta P_{n}^{(2\beta,2\alpha)}(iX/\sqrt{q}),
\end{equation}
where $X(u)=\sinh_{q}(u/r)$.  Note that the solutions have the following properties,
\bea
P_{n}^{(2\alpha,2\beta)*}(Y)&=&(-1)^{n}P_{n}^{(\pm2\alpha,\pm2\beta)}(Y),\quad \text{for} \quad \alpha=\pm\beta^{*}.
\eea
For $\alpha=\beta^{*}$, the Jacobi polynomial is real~(imaginary) for even~(odd) $n$.  Notably for this case, the spatial wave function $\Big(\sqrt{q}+iX\Big)^\alpha \Big(\sqrt{q}-iX\Big)^\beta P_{n}^{(2\beta,2\alpha)}(iX/\sqrt{q})$ for even $n$ is also real and the energy given by (\ref{E01}) depends only on the quantum number $n$.  This energy is independent of the spin-orbit and magnetic field.

\subsubsection{Asymptotic solutions}  \label{asymp1}

In this section, we explore the asymptotic solutions at the wormhole throat $u=0$ and the Hilbert horizon $\displaystyle{u=u_{H}=r\log\Big(\frac{r}{a}+\sqrt{q+\frac{r^{2}}{a^{2}}}\Big)}$.  Generically, linear differential equation of the form
\be
f(Y)\Phi''(Y)+g(Y)\Phi'(Y)+h(Y)\Phi(Y)=0,
\ee
has asymptotic solutions in the region around $u=u_{0}$, where $f(Y)\simeq f(Y(u_{0}))\equiv f_{0}, g(Y)\simeq g(Y(u_{0}))\equiv g_{0}, h(Y)\simeq h(Y(u_{0}))\equiv h_{0}$, in the exponential form 
\be
\Phi(Y)=\Phi_{0}e^{\delta Y},
\ee
$Y=Y(u)$ and
\be
\delta = \frac{1}{2f_{0}}\Big(-g_{0}\pm \sqrt{g_{0}^{2}-4 f_{0}h_{0}}\Big).
\ee
Around the throat $u\simeq 0$, Eq.~(\ref{eq:34}) gives $f_{0}=1, g_{0}=2(\alpha - \beta), h_{0}=-(\alpha+\beta)(\alpha+\beta+1)-k^{2}r^{2}-1/4$ and thus
\be
\delta = -(\alpha - \beta)\pm \sqrt{(\alpha -\beta)^{2}-2n(\alpha+\beta)-n(n+1)}.
\ee
Around the Hilbert horizon $u\simeq u_{H}, Y_{H}=Y(u_{H})$, Eq.~(\ref{eq:34}) gives 
\be
f(Y_{H})=1+\frac{r^{2}}{q a^{2}}, g(Y_{H})=2\Big(\alpha - \beta -(\alpha +\beta +1)\frac{ir}{a\sqrt{q}}\Big), h(Y_{H})=-(\alpha+\beta)(\alpha+\beta+1)-k^{2}r^{2}-\frac{1}{4}
\ee
and 
\be
\delta = \frac{1}{1+r^{2}/qa^{2}}\Big[-(\alpha -\beta )+\frac{i r (\alpha +\beta +1)}{a \sqrt{q}}\pm\sqrt{-\frac{n \left(a^2 q+r^2\right) (2 \alpha +2 \beta +n+1)}{a^2 q}+\left(\frac{i r (\alpha +\beta +1)}{a \sqrt{q}}-\alpha +\beta \right)^2}\Big].
\ee
 Since $Y=i\sinh_{q}(\displaystyle{\frac{u}{r}})/\sqrt{q}=i\displaystyle{\sqrt{\frac{R^{2}}{qa^{2}}-1}}~(q \leq 1, R>a)$ and $\Phi \sim e^{\delta Y}$, sign of the real part of $\delta$ determines the direction of traveling waves in the asymptotic regions, positive sign~($\text{Re}~\delta >0$) implies waves traveling in $+u$ or $+R$ direction and vice versa.  We can substitute $\alpha, \beta$ and obtain complicated expressions of $(\text{Re}~\delta, \text{Im}~\delta)$ depending on the values of $n, m', \sigma, q, a, r$.  Alternatively, the momentum at each boundary can be determined by the operator $-i\partial_{u}$ and $-i\partial_{R}$ at $u=0$ and $R=R(u_{H})\equiv R_{H}=\sqrt{r^{2}+q a^{2}}$ respectively. The imaginary part of the momentum~(or $\delta$) can be interpreted as the {\it momentum dissipation} along $u$ direction.
 
All possible solutions and their asymptotic behaviours are summarized in Table~\ref{ConFlux}. The four choices $(+,+),(-,-),(+,-),(-,+)$ of $(\alpha,\beta)$ correspond to $(\kappa_{1},\kappa_{2})=(+1,+1), (-1,-1),(+1,-1),(-1,+1)$ where
\begin{eqnarray}
\alpha &=& \kappa_{1}\Big(\frac{1}{4}+\frac{i}{\sqrt{q}}\frac{\sigma m'r}{2a}\Big), \beta = \kappa_{2}\Big(\frac{1}{4}-\frac{i}{\sqrt{q}}\frac{\sigma m'r}{2a}\Big)
\end{eqnarray}
respectively. The real part $\text{Re}~\delta$ depends on choices of $\alpha, \beta$ and it could be either positive or negative. Wave at the boundary is either outgoing or incoming depending on the value of $n,m'$ and $q$. For $1\geq q > 0, n \geq 0$, solution $(+,+)$ and $(-,-)$ are purely dissipative or unstable at both the throat $u=0$ and the Hilbert horizon $u=u_{H}$ since $\text{Re}~\delta=0$. The choice of sign of the square-root term of $\delta$ determines whether we choose attenuating~(i.e., decreasing wave function, $\text{Im}~\delta>0$) or enhancing~(i.e., increasing wave function, $\text{Im}~\delta<0$) mode.  Both modes are useful for describing scattering phenomenon where incoming~(outgoing) waves from outside~(inside) region scatter with the wormhole interior~(exterior).  This problem is currently being investigated~\cite{BurikhamF}.  

On the other hand, solution $(+.-)$ and $(-,+)$ generally have nonzero $\text{Re}~\delta$ and $\text{Im}~\delta$ at $u=0, u_{H}$.  Again, the sign choice of the square-root term determines whether the wave is outgoing or incoming at both boundaries.  Incoming~(outgoing) waves are suitable when describe the scattering of waves from the outer~(inner) region of the wormhole.  

\begin{table}[h!]
\begin{tabular}{ |c|c|c|c|c| } 
 \hline
 $\alpha$ & $\beta$ & $E_{n,m}^{2}-M^{2}c^{4}$ & $\delta(u=0)$ & $\delta(u_{H})\left(1+r^{2}/qa^{2}\right)$ \\ 
 
 \hline
 
 &  &  &   & \\ 

 $+$ & $+$ & $-\frac{\hbar^{2}c^{2}}{r^2}\L(n+1\R)^{2}$ & $\pm i\sqrt{\frac{m'^2 r^2 \sigma ^2}{a^2 q}+n (n+2)}-i\frac{m' r \sigma }{a \sqrt{q}}$ & $\pm \frac{1}{2} i\sqrt{\frac{4 a^2 (n+2) n q+r^2 \left((3-2 m' \sigma )^2+4 n^2+8 n\right)}{a^2 q}}+i\frac{r\left(\frac{3}{2}-m' \sigma \right)}{a \sqrt{q}}$ \\  
 
  &  &  &  & \\ 
  
 \hline
 
  &  &  &  & \\ 
  
 $-$ & $-$ & $-\frac{\hbar^{2}c^{2}}{r^2}n^{2}$  &  $\pm i\sqrt{\frac{m'^2 r^2 \sigma ^2}{a^2 q}+n^2}+i\frac{m' r \sigma }{a \sqrt{q}}$ & $\pm\frac{1}{2} i\sqrt{\frac{4 a^2 n^2 q+r^2 \left((2 m' \sigma +1)^2+4 n^2\right)}{a^2 q}}+i\frac{r\left(m' \sigma +\frac{1}{2}\right)}{a \sqrt{q}}$\\ 
 
  &  &  &  & \\ 
  
 \hline
 
  &  &  &  &  \\ 
  
  $+$ & $-$ & $-\frac{\hbar^{2}c^{2}}{r^2}\L[n+\frac{1}{2}+i\frac{\sigma m'r}{a\sqrt{q}}\R]^{2}$ &  $-\frac{1}{2}\pm\frac{1}{2}\sqrt{-\frac{8 i m' n r \sigma }{a \sqrt{q}}-4 n^2-4n+1}$ & $\pm\frac{1}{4} \sqrt{\frac{\left(2 a^2 q-4 i a \sqrt{q} r+4 m' r^2 \sigma \right)^2-8 a n \sqrt{q} \left(a^2 q+r^2\right) \left(2 a (n+1) \sqrt{q}+4 i m' r \sigma \right)}{a^4 q^2}}$ \\
 &&&& $-\frac{1}{2}-\frac{m' r^2 \sigma }{a^2 q}+i\frac{r}{a \sqrt{q}}$  \\ 
  
  &  &  &  &   \\ 
   
 \hline
 
  &  &  &  &  \\ 
  
  $-$ & $+$ & $-\frac{\hbar^{2}c^{2}}{r^2}\L[n+\frac{1}{2}-i\frac{\sigma m'r}{a\sqrt{q}}\R]^{2}$  &  $\frac{1}{2}\pm\frac{1}{2}\sqrt{\frac{8 i m' n r \sigma }{a \sqrt{q}}-4 n^2-4n+1}$ & $\pm\frac{1}{4} \sqrt{\frac{4 \left(a^2 q+2 i a \sqrt{q} r+2 m' r^2 \sigma \right)^2-8 a n \sqrt{q} \left(a^2 q+r^2\right) \left(2 a (n+1) \sqrt{q}-4 i m' r \sigma \right)}{a^4 q^2}}$ \\
 &&&& $+\frac{1}{2}+\frac{m' r^2 \sigma }{a^2 q}+i\frac{r}{a \sqrt{q}}$ \\ 
  
   &  &  &  & \\ 
   
 \hline
\end{tabular}
\caption{Summary of the energy levels $E_{n,m'}$, and the asymptotic parameter $\delta$ of the solutions in $(1+2)$-dimensional wormhole with constant magnetic flux. } \label{ConFlux}
\end{table}

Emergence of QNMs in this $(1+2)$ dimensional wormhole should be compared with the situation in astrophysical or gravitational traversable wormhole in $(1+3)$ dimensions.  In Ref.~\cite{Konoplya2005,Konoplya2010, Konoplya2018}, ringing of astrophysical wormhole results in the QNMs due to the leaking-out waves into the asymptotically flat infinity and the throat~(and subsequently to the other asymptotically flat region).  The traversable $(1+3)$-dimensional wormhole has no event horizon and is the analog of our $(1+2)$-dimensional ``wormhole''~(minus the time dilatation in the latter).  Choosing only the leaking-out boundary condition leads to singling out only the decaying QNMs.  In our case, we keep all possible boundary conditions in this work since they are useful in the generic scattering processes.

\subsection{Constant magnetic field solution}

Another physical situation that can give insight to the role of magnetic field on the fermion is the uniform magnetic field environment.  We will show that again the spin-orbit coupling is induced by the curvature of space or ``gravity'' of the wormhole.  Unstable modes and QNMs will be generated again from such interaction.  However, there is additional momentum-diffusive modes~(negative $p^{2}_{u}$) that also have dependency on the coupling between orbital angular momentum and the magnetic field.  This leads to a new $p_{u}$-diffusive modes that depend on the spin of the fermion in the wormhole.  The $p_{u}$-diffusive modes can have either real or imaginary energy depending on the quantum number $n$ and the magnetic field in comparison to the rest-mass energy $Mc^{2}$.

For constant uniform magnetic field, the equation of motion is modified since now the operator $\textbf{D}^2$ is given by
\begin{equation}\label{Dsq1}
\begin{split}
\textbf{D}^2
          &=\sigma^{1}\sigma^{1}\Big(\partial_{u}+ \frac{R'}{2R}\Big)^2+ \sigma^{2}\sigma^{2}\Big(\frac{1}{R}\partial_{v}-\frac{ie}{2\hbar c}BR\Big)^2   \\
          &\quad+\sigma^{1}\sigma^{2}\Big(\partial_{u}+ \frac{R'}{2R}\Big)\Big(\frac{1}{R}\partial_{v}-\frac{ie}{2\hbar c}BR\Big)+\sigma^{2}\sigma^{1}(\frac{1}{R}\partial_{v}-\frac{ie}{2\hbar c}BR\Big)\Big(\partial_{u}+ \frac{R'}{2R}\Big)   \\
          &=\partial_{u}^2+\frac{1}{R^2}\partial_{v}^2
          +\frac{R'}{R}\partial_{u}
          -\Big(i\sigma^{3}\frac{R'}{R^2}+\frac{ie}{\hbar c}B\Big)\partial_{v}
          +\frac{1}{2}\Big(\frac{R''}{R}\Big)
          -\Big(\frac{R'}{2R}\Big)^2
          +\Big(\frac{ie}{2\hbar c}BR\Big)^2
          -i\sigma^{3}\frac{ie}{2\hbar c}BR'
\end{split}
\end{equation}  
With the same separation of variables (\ref{varsep}), we obtain
\begin{equation}\label{eomB0}
\begin{split}
0
          &=\Big[\partial_{u}^2+\frac{R'}{R}\partial_{u}
          +k^2+\frac{eB}{\hbar c}m
          +\frac{R'\sigma m -m^2 -\Big(\frac{R'}{2}\Big)^2 }{R^2}
          +\frac{1}{2}\Big(\frac{R''}{R}\Big)
          -\Big(\frac{eB}{2\hbar c}\Big)^{2} R^2
          +\frac{eB}{2\hbar c}\sigma R'
          \Big]\varphi^{\sigma}_{II}(u).
\end{split}
\end{equation} 
Again, a shape of deformed hyperbolic wormhole $R(u)=a\cosh_{q}(u/r)$ is chosen for further calculation. After performing the similar transformations using $X(u)=\sinh_{q}(u/r)$ and $\varphi(X)=(\sqrt{q}+iX)^{\alpha'} (\sqrt{q}-iX)^{\beta'}\Phi(X)$, the equation of motion becomes
\begin{equation}\label{eomB}
\begin{split}
0
         &=(q+X^2)\Phi''(X) +\Big[\textup{A}+\textup{B}X\Big]\Phi'(X)
         +\Big[\textup{C}+\textup{D}X+\textup{E}X^2+k^2 r^2 \Big]\Phi(X),
\end{split}
\end{equation}
where we again assume
\begin{equation}\label{eq:30}
2i\Big(\alpha'^2-\beta'^2\Big)\sqrt{q}+\frac{r}{a}\sigma m=0, 
\qquad -2\Big(\alpha'^2+\beta'^2\Big)q+\frac{q}{4}-\Big(\frac{r}{a}m\Big)^2=0,  \nonumber
\ee
leading to
\be
\alpha'^2=\Big(\frac{1}{4}+\frac{i}{\sqrt{q}}\frac{\sigma mr}{2a}\Big)^2, 
\qquad \beta'^2=\Big(\frac{1}{4}-\frac{i}{\sqrt{q}}\frac{\sigma mr}{2a}\Big)^2.
\end{equation}
The coefficient parameters are defined as the following
\begin{equation}\label{eq:31}
\begin{split}
\textup{A}
         &=2i(\alpha'-\beta')\sqrt[]{q},\quad \textup{B}
         =2(\alpha'+\beta'+1),\quad\textup{C}
         =(\alpha'+\beta')(\alpha'+\beta'+1)+\frac{1}{4}+\frac{eB}{\hbar c}mr^2-\Big(\frac{ar}{2}\frac{eB}{\hbar c}\Big)^{2}q,                            \\
\textup{D}
         &=\frac{ar}{2}\frac{eB}{\hbar c}\sigma,\quad
\textup{E}
         =-\Big(\frac{ar}{2}\frac{eB}{\hbar c}\Big)^{2}.
\end{split}
\end{equation}

In order to find the solution to Eq.~(\ref{eomB}), we first obtain the asymptotic solution for large $X$, Eq.~(\ref{eomB}) now becomes
\begin{equation}\label{eq:42}
0\approx X^2\Phi_0''(X)+\textup{E}X^2\Phi_0(X),
\end{equation}
having the solutions: $\Phi_0(X)=\exp[\pm\sqrt{-\textup{E}}X]=\exp[\pm\frac{ar}{2}\frac{eB}{\hbar c}X]$.  Rewriting the solution for all region as
\begin{equation}\label{eq:43}
\Phi(X)=\tilde{\Phi}(X)\Phi_0(X),
\end{equation}
the equation of motion (\ref{eomB}) then takes the form
\begin{equation}\label{eq:44}
0=(q+X^2)\tilde{\Phi}''(X)
         +\Big[\text{F}+\textup{G}X+\text{H}X^2\Big] \tilde{\Phi}'(X)+\Big[\text{I}+k^2 r^2 +\text{J}X\Big]\tilde{\Phi}(X),
\end{equation}
where the parameters are defined as
\begin{equation}\label{eq:45}
\begin{split}
\textup{F}
         &=\textup{A}\pm2q\sqrt{-\textup{E}},\quad
\textup{G}
         =\textup{B},\quad \textup{H}
         =\pm2\sqrt{-\textup{E}}=\pm ar\frac{eB}{\hbar c}                    \\
\textup{I}
         &=\textup{C}-q\textup{E}\pm\sqrt{-\textup{E}}\textup{A},\quad \textup{J}
         =\textup{D}\pm\textup{B}\sqrt{-\textup{E}}.
\end{split}
\end{equation}
For $\alpha'=\beta'^{*}=\displaystyle{\frac{1}{4}+\frac{i\sigma m r}{2a\sqrt{q}}}$,  the parameters are explicitly
\bea
\text{A}&=&-2\frac{\sigma m r}{a}, \quad\text{B}=\text{G}=3, \quad\text{C}=1+\frac{eB}{\hbar c}mr^{2}-q\left( \frac{areB}{2\hbar c}\right)^{2}, \\
\text{F}&=&-2\frac{\sigma m r}{a}\pm qar\frac{eB}{\hbar c}, \quad\text{I}=1+(1\mp\sigma)mr^{2}\frac{eB}{\hbar c}, \quad\text{J}=(\sigma \pm 3)\frac{areB}{2\hbar c}, 
\eea
and for $\alpha'=-\beta'^{*}=\displaystyle{\frac{1}{4}+\frac{i\sigma m r}{2a\sqrt{q}}}$,
\bea
\text{A}&=&i\sqrt{q}, \quad\text{B}=\text{G}=2\left( 1+\frac{i\sigma m r}{a\sqrt{q}}\right), \quad\text{C}=\frac{1}{4}+\frac{eB}{\hbar c}mr^{2}-q\left( \frac{areB}{2\hbar c}\right)^{2}+\frac{i\sigma m r}{a\sqrt{q}}-\frac{(mr)^{2}}{qa^{2}}, \\
\text{F}&=&i\sqrt{q}\pm \frac{areB}{\hbar c}, \quad\text{I}=\left( \frac{1}{2}+\frac{i\sigma m r}{a\sqrt{q}}\right)^{2}+mr^{2}\frac{eB}{\hbar c}\pm \frac{iareB\sqrt{q}}{2\hbar c}, \quad\text{J}=(\sigma \pm 2)\frac{areB}{2\hbar c}\pm \frac{i\sigma m r^{2}eB}{\hbar c \sqrt{q}},
\eea
respectively.

\subsubsection{Small X approximation}

The wave function can be solved exactly for small $X$~($\sinh_{q}(u/r)=X<1$ implies that $r<a$ since $X_{H}\equiv X(u_{H})=r/a$) in terms of the Jacobi polynomials as we will show in the following.  For small $X$, Eq.~(\ref{eq:44}) takes the form
\begin{equation}\label{eq:51}
0\approx(q+X^2)\tilde{\Phi}''(X)
         +\Big[\textup{F}+\textup{G}X\Big] \tilde{\Phi}'(X)
         +\Big[\textup{I}+k^2 r^2\Big]\tilde{\Phi}(X).                                                                   \ee
Changing variable $X=-i\sqrt[]{q}Y$, the equation becomes
\be
0=(1-Y^2)\tilde{\Phi}''(Y)
         +\Big[-i\frac{\textup{F}}{\sqrt{q}}-\textup{G}Y\Big] \tilde{\Phi}'(Y)-\Big[\textup{I}+k^2 r^2\Big]\tilde{\Phi}(Y).
\end{equation}
For the choice $\Phi_0(X)=\exp[-\sqrt{-\textup{E}}X]$, Eq.~(\ref{eq:51}) has solution in the form of the Jacobi polynomials, given conditions
\begin{equation}\label{eq:52}
\beta''-\alpha''=-i\frac{\textup{F}}{\sqrt{q}}, \qquad \beta''+\alpha''+2=\textup{G}, \qquad n\Big[n+\beta''+\alpha''+1\Big]=-\Big[\textup{I}+k^2 r^2 \Big],                
\ee
therefore
\be
\begin{split}
\beta''
        &=\frac{1}{2}\Big(\textup{G}-2-i\frac{\textup{F}}{\sqrt[]{q}}\Big), \alpha''=\frac{1}{2}\Big(\textup{G}-2+i\frac{\textup{F}}{\sqrt[]{q}}\Big),              \\
-k^{2}_{n} r^2
        &=I+n\Big(n+\alpha''+\beta''+1\Big).
\end{split}
\end{equation}
And the solution is 
\begin{equation}\label{eq:53}
\tilde{\Phi}(Y)=P^{(\alpha'',\beta'')}_{n}\Big(Y\Big).
\end{equation}
The quantization of energy for $\alpha'=\beta'^{*}$ case is
\begin{equation}\label{eq:54}
E^{2}_{n}-M^2c^4=k^{2}_{n} \hbar^2 c^2=-\frac{\hbar^2 c^2}{r^2}\Big[(n+1)^{2}+(1+\sigma) mr^2 \frac{eB}{\hbar c}\Big].
\end{equation}
For the other possibility $\alpha'=-\beta'^{*}$, the energy is
\begin{equation}\label{econB2}
E^{2}_{n}-M^2c^4=k^{2}_{n} \hbar^2 c^2=-\frac{\hbar^2 c^2}{r^2}\Big[\left(n+\frac{1}{2}+\frac{i\sigma m r}{a\sqrt{q}}\right)^{2}+mr^2 \frac{eB}{\hbar c} - \frac{iareB\sqrt{q}}{2\hbar c }\Big].
\end{equation}
The energy given by Eq.~(\ref{eq:54}) has an energy splitting between the spin up~($\sigma =1$) and down~($\sigma = -1$) proportional to $2mr^{2}eB/\hbar c$.  This is the spin-orbit-magnetic coupling.  Notably, the spin-down state does not feel the magnetic field.  For sufficiently large $n, m, B$, the energy becomes purely imaginary since the negative interaction energy is larger than the rest-mass energy $Mc^{2}$.  On the other hand, the energy given by Eq.~(\ref{econB2}) is complex with the imaginary part depending on both the spin-orbit and the external magnetic field.  QNMs always exist for nonzero $m$ and magnetic field in this case.    

For $(\alpha',\beta')\to (-\alpha',-\beta')$ cases, the energies become
\begin{equation}\label{eq:54n}
E^{2}_{n}-M^2c^4=k^{2}_{n} \hbar^2 c^2=-\frac{\hbar^2 c^2}{r^2}\Big[n^{2}+(1-\sigma) mr^2 \frac{eB}{\hbar c}\Big]
\end{equation}
for $\alpha'=\beta'^{*}=-\displaystyle{\frac{1}{4}-\frac{i\sigma m r}{2a\sqrt{q}}}$ and
\begin{equation}\label{econB2n}
E^{2}_{n}-M^2c^4=k^{2}_{n} \hbar^2 c^2=-\frac{\hbar^2 c^2}{r^2}\Big[\left(n+\frac{1}{2}-\frac{i\sigma m r}{a\sqrt{q}}\right)^{2}+mr^2 \frac{eB}{\hbar c} + \frac{iareB\sqrt{q}}{2\hbar c }\Big]
\end{equation} 
for $\alpha'=-\beta'^{*}=-\displaystyle{\frac{1}{4}-\frac{i\sigma m r}{2a\sqrt{q}}}$ respectively.

\subsubsection{Asymptotic solutions}

Similar to the constant-flux scenario in Section \ref{asymp1}, the same analysis can be performed to explore the behaviour of solutions at both boundaries $u=0, u_{H}$.  We can express the equation of motion (\ref{eq:44}) in the asymptotic region in the form  
\begin{equation}\label{eq:FieldNearZero01}
\begin{split}
0=&A_{0,H}\tilde{\Phi}''(Y)+B_{0,H}\tilde{\Phi}'(Y)+C_{0,H}\tilde{\Phi}(Y),
\end{split}
\end{equation}
where we define at $u_{0}$
\begin{equation}\label{eq:FieldNearZero02}
\begin{split}
A_{0}\equiv& \/ 1   \\
B_{0}\equiv& \/ 2(\alpha'-\beta')+i\sqrt{q}\frac{ar}{L^{2}}  \\
C_{0}\equiv& \/ n\Big( n+1+2(\alpha' + \beta')\Big),
\end{split}
\end{equation} 
and
\begin{equation}\label{eq:FieldNearuH02}
\begin{split}
A_{H}\equiv& \/ 1+\frac{r^{2}}{qa^{2}}   \\
B_{H}\equiv& \/ 2(\alpha'-\beta')+\frac{ir}{\sqrt{q}a}\L[\frac{r^{2}+qa^{2}}{L^{2}}-2(\alpha'+\beta'+1)\R]   \\
C_{H}\equiv& \/ n\Big( n+1+2(\alpha' + \beta')\Big)-\frac{r^{2}}{L^{2}}\Big( \frac{\sigma}{2}-(1+\alpha' +\beta') \Big)
\end{split}
\end{equation} 
at $u_{H}$.  The corresponding asymptotic parameters are then given by
\begin{equation}\label{eq:FieldAsymp}
\delta_{0,H}=-\frac{B_{0,H}}{2A_{0,H}}\pm\sqrt{\L(\frac{B_{0,H}}{2A_{0,H}}\R)^{2}-\frac{C_{0,H}}{A_{0,H}}}.
\end{equation}
They are summarized in Table \ref{Tabel:NearZero} and \ref{Tabel:NearuH}.  

\begin{table}[h!]
\centering
\begin{tabular}{ |c|c|c|c| } 
 \hline
 $\alpha'$ & $\beta'$ & $E^{2}-M^{2}c^{4}$ & $\delta_{0}$ \\ 
 
 \hline
 
 &  &  &     \\ 

 $+$ & $+$ & $-\hbar^{2}c^{2}\displaystyle{\L[(n+1)^{2}+(1+\sigma)m\frac{r^{2}}{L^{2}}\R]}$ & $\displaystyle{-i\L(\sqrt{q}\frac{a r}{2L^{2}}+\frac{r}{a}
 \frac{m\sigma}{\sqrt{q}}\R)\pm i\sqrt{n(n+2)+\L(\sqrt{q}\frac{a r}{2L^{2}}+\frac{r}{a}
 \frac{m\sigma}{\sqrt{q}}\R)^{2}}}$ \\  
 
  &  &  &     \\ 
  
 \hline
 
  &  &  &     \\ 
  
 $-$ & $-$ & $-\hbar^{2}c^{2}\displaystyle{\L[n^{2}+(1-\sigma)m\frac{r^{2}}{L^{2}}\R]}$  &  $\displaystyle{-i\L(\sqrt{q}\frac{a r}{2L^{2}}-\frac{r}{a}
 \frac{m\sigma}{\sqrt{q}}\R)\pm i\sqrt{n^{2}+\L(\sqrt{q}\frac{a r}{2L^{2}}-\frac{r}{a}
 \frac{m\sigma}{\sqrt{q}}\R)^{2}}}$\\ 
 
  &  &  &    \\ 
  
 \hline
 
  &  &  &    \\ 
  
  $+$ & $-$ & $-\hbar^{2}c^{2}\displaystyle{\L[\L(n+\frac{1}{2}+\frac{irm\sigma}{a\sqrt{q}}\R)^{2}+\L(m-i\sqrt{q}\frac{a}{2r}\R)\frac{r^{2}}{L^{2}}\R]}$ &    $\displaystyle{-\L(i\sqrt{q}\frac{a r}{2L^{2}}+\frac{1}{2}\R)\pm i\sqrt{n^{2}+n-\frac{1}{4}+\frac{q}{4}\frac{a^{2} r^{2}}{L^{4}}-i\L(\sqrt{q}\frac{a r}{2L^{2}}-2\frac{mn\sigma r}{\sqrt{q}a}\R)}}$    \\ 
  
  &  &  &    \\ 
   
 \hline
 
  &  &  &     \\ 
  
  $-$ & $+$ & $-\hbar^{2}c^{2}\displaystyle{\L[\L(n+\frac{1}{2}-\frac{irm\sigma}{a\sqrt{q}}\R)^{2}+\L(m+i\sqrt{q}\frac{a}{2r}\R)\frac{r^{2}}{L^{2}}\R]}$ &    $\displaystyle{-\L(i\sqrt{q}\frac{a r}{2L^{2}}-\frac{1}{2}\R)\pm i\sqrt{n^{2}+n-\frac{1}{4}+\frac{q}{4}\frac{a^{2} r^{2}}{L^{4}}+i\L(\sqrt{q}\frac{a r}{2L^{2}}-2\frac{mn\sigma r}{\sqrt{q}a}\R)}}$   \\ 
  
   &  &  &   \\ 
   
 \hline
\end{tabular}
\caption{The asymptotic parameter $\delta_{0}$ at $u=0$ for constant field scenario.} \label{Tabel:NearZero}
\end{table}

\begin{table}[H]
\centering
\begin{tabular}{|c|c|c|c|}
 \hline
 $\alpha'$ & $\beta'$ & $\delta_{H}A_{H}$ \\ 
 
 \hline
 
     &  &      \\ 
 
$+$  & $+$ &  $\displaystyle{-\frac{ir}{a\sqrt{q}}\L(\frac{r^{2}+qa^{2}}{2L^{2}}-\frac{3}{2}+m\sigma\R)\pm i\sqrt{\frac{r^{2}+qa^{2}}{qa^{2}}\L(n(n+2)+(3-\sigma)\frac{r^{2}}{2L^{2}}\R)+\frac{r^{2}}{qa^{2}}\L(\frac{r^{2}+qa^{2}}{2L^{2}}-\frac{3}{2}+m\sigma\R)^{2}}}$    \\

     &  &      \\ 

 \hline
 
     &  &      \\ 
   
$-$  & $-$ &  $\displaystyle{-\frac{ir}{a\sqrt{q}}\L(\frac{r^{2}+qa^{2}}{2L^{2}}-\frac{1}{2}-m\sigma\R)\pm i\sqrt{\frac{r^{2}+qa^{2}}{qa^{2}}\L(n^{2}+(1-\sigma)\frac{r^{2}}{2L^{2}}\R)+\frac{r^{2}}{qa^{2}}\L(\frac{r^{2}+qa^{2}}{2L^{2}}-\frac{1}{2}-m\sigma\R)^{2}}}$  \\ 

     &  &      \\ 
  
 \hline
 
     &  &     \\ 
    
 $+$ & $-$ &  $\displaystyle{\pm i\sqrt{\frac{r^{2}+qa^{2}}{qa^{2}}\L(n^{2}+n+2\frac{ir}{\sqrt{q}a}mn\sigma+\frac{r^{2}}{L^{2}}\L(1-\frac{\sigma}{2}+\frac{ir}{\sqrt{q}a}m\sigma\R)\R)-\L(\frac{1}{2}+\frac{r^{2}}{qa^{2}}m\sigma+\frac{ir}{a\sqrt{q}}\L(\frac{r^{2}+qa^{2}}{2L^{2}}-1\R)\R)^{2}}}$   \\ [0.3cm]
  
     &  &    $\displaystyle{-\L(\frac{1}{2}+\frac{r^{2}}{qa^{2}}m\sigma\R)-\frac{ir}{a\sqrt{q}}\L(\frac{r^{2}+qa^{2}}{2L^{2}}-1\R)}$     \\ 

     &  &      \\ 
   
 \hline
 
     &  &      \\ 
    
 $-$ & $+$ &  $\displaystyle{\pm i\sqrt{\frac{r^{2}+qa^{2}}{qa^{2}}\L(n^{2}+n-2\frac{ir}{\sqrt{q}a}mn\sigma+\frac{r^{2}}{L^{2}}\L(1-\frac{\sigma}{2}-\frac{ir}{\sqrt{q}a}m\sigma\R)\R)-\L(\frac{1}{2}+\frac{r^{2}}{qa^{2}}m\sigma-\frac{ir}{a\sqrt{q}}\L(\frac{r^{2}+qa^{2}}{2L^{2}}-1\R)\R)^{2}}}$  \\ [0.3cm]
   
     &  &   $\displaystyle{+\L(\frac{1}{2}+\frac{r^{2}}{qa^{2}}m\sigma\R)-\frac{ir}{a\sqrt{q}}\L(\frac{r^{2}+qa^{2}}{2L^{2}}-1\R)}$    \\ 

     &  &      \\ 
   
 \hline
 \end{tabular}
\caption{The asymptotic parameter $\delta_{H}$ expressed in product with $A_{H}$ at $u_{H}$ for constant field scenario.}\label{Tabel:NearuH}
\end{table}
The states given by combination $(\alpha,\beta)=(+,+), (-,-)$ give purely momentum dissipative modes at both boundaries, similar behaviour to the constant flux scenario.  The other states from $(\alpha,\beta)=(+,-), (-,+)$ combinations have nonzero $\text{Re}~\delta$, representing traveling incoming and outgoing waves with attenuation or enhancement depending on $\text{Im}~\delta$. All possibilities are involved in consideration of the scattering processes which we leave as the future work.  

\subsection{Cylindrical Wormhole}  \label{conR}

To understand essential physics of the magnetized charged fermion in the wormhole, consider a simple case when $R(u)$ is constant, i.e. a wormhole tube.  In this case, the intrinsic~(or Gaussian) curvature of the wormhole is zero so we can identify which effects are induced by the wormhole ``gravity''.  The two separate magnetic scenarios of constant flux and field reduce to the same physical system as both (\ref{eq:25}) and (\ref{eomB}) become~(suppressing subscript and superscript)
\begin{equation}\label{eq:55}
\begin{split}
0
          &=\Big[\partial_{u}^2
          -\Big(\frac{m'}{R}\Big)^2 +k^2
          \Big]\varphi(u)=\Big[\partial_{u}^2+k^2+\frac{eB}{\hbar c}m-\frac{m^2}{R^2}          -\Big(\frac{eB}{2\hbar c}\Big)^{2} R^2 \Big]\varphi(u).
\end{split}
\end{equation}
Assuming the solution in the form  $\varphi_{m}(u)=\varphi_{m}(0)\exp({i\texttt{k}_u u})$ to obtain
\begin{equation}\label{eq:56}
R^{2}k^{2}=R^{2}\frac{E^{2}_{m}(\texttt{k}_{u})-M^2c^4}{\hbar^2 c^2}
          =\Big(R\texttt{k}_{u}\Big)^2+\left(m-\frac{\phi}{\phi_{0}}\right)^2.
\end{equation} \\
Setting $2r$ as the length of the cylinder with the boundary conditions $\varphi(u=0)=\varphi(u=2r)=0$, we get $\texttt{k}_{u}=n\pi/2r$ where $n=0,1,2,...$ The quantization of energy via Eq.~(\ref{eq:56}) is then
\begin{equation}\label{eq:57}
E^{2}_{m,n}=M^2c^4+\Big(\frac{\hbar c}{R}\Big)^{2}\Big[\Big(\frac{n\pi R}{2r}\Big)^2+\left(m-\frac{\phi}{\phi_{0}}\right)^2\Big]=M^2c^4+\Big(\frac{\hbar c}{R}\Big)^{2}\Big[(\texttt{k}_{u}R)^2+\left( m-\frac{eBR^{2}}{2\hbar c}\right)^{2}\Big].
\end{equation}
The energy is purely real and only normal modes exist.  The spin-orbit coupling disappears together with the Landau coupling between spin and the magnetic field.  The only remaining interaction is the orbital-magnetic Landau coupling.

As discussed in Ref.~\cite{Wilczek}, the fermions can have arbitrary orbital angular momentum which results in various possible statistics.  For $\phi/\phi_{0}=\text{integer}+1/2$, the fermion statistics become Bose-Einstein~(i.e., like bosons) in the magnetized wormhole.  They can condensate and flow like superfluid along the hole.   

\section{Beltrami and Elliptic pseudosphere wormhole }  \label{BelEll}

There are special cases when the deformation parameter $q=0, -1$, i.e. the Beltrami and elliptic pseudosphere wormhole that are not captured in the general analysis, we address them in this section. 

\subsection{Beltrami wormhole}

First we consider the {\it constant flux} scenario.  For ${q=0}$ and ${R(u)=\frac{a}{2} \e^{u/r}}$ of the Beltrami wormhole, Eq.(\ref{eq:25}) become
\begin{equation}\label{eq:Beltrami1}
\begin{split}
0
          &=\L[\partial_{u}^2
          +\frac{1}{r}\partial_{u}
          +\frac{1}{2r^{2}}-\L(\frac{1}{2r}\R)^2
          +2\frac{m'\sigma}{ar}\e^{-u/r}-\L(2\frac{m'}{a}\R)^2\e^{-2u/r}+k^2\R] \varphi(u).
\end{split}
\end{equation}
The general solution can be expressed in the form
\begin{equation}\label{eq:Beltrami8}
\varphi(u)=e^{-Z/2}Z^{- i k r+\frac{1}{2}}\L[C_{1}~{ }_{1}F_{1}\L(-i k r+\frac{1-\sigma}{2},1-2ikr, Z\R)+C_{2}U\L(-i k r+\frac{1-\sigma}{2},1-2ikr, Z\R)\R],
\end{equation}
where $Z=\displaystyle{\frac{4 m' r e^{-u/r}}{a}}$ that takes the value $Z=4m'r/a, 0$ for $u=0, \infty$ respectively.  $U(a,b,Z)$ is the confluent hypergeometric function of the second kind.

Stationary-states solutions require finite wave functions at $Z=4m'r/a, 0$ even though the wormhole actually ends at the Hilbert horizon $u_{H}=r\log(2r/a), R_{H}=r, Z_{H}=2m'$~(where $R'=1$).  Regularity at $Z=4m'r/a > 1$ demands that the series of the hypergeometric function truncates at finite power of $Z$ giving
\be
\mp i k r+\frac{1-\sigma}{2}=-n,
\ee
for non-negative integer $n$.  This leads to the energy quantization
\be
E^{2}_{n}-M^2c^4=k^{2}_{n} \hbar^2 c^2=-\frac{\hbar^2 c^2}{r^2}\left[ n+\Big( \frac{1-\sigma}{2}\Big) \right]^{2}.  
\ee
Remarkably, the energies do not depend on $B$ and $m$ at all, only the wave functions have $m'$ dependence.  All $m'$ states degenerate in each energy level $E_{n}$.

A special solution for $m'=0$ where the magnetic flux is quantized to integer values $\phi/\phi_{0}=m=0,1,2,...$ can be obtained from (\ref{eq:25}), giving
\be
\phi(u)=C_1 e^{-u/2r} e^{-i ku}+C_2 e^{-u/2r} e^{i ku}.
\ee
The solutions are decaying plane wave traveling in the $u$ direction, in and out of the wormhole.  The wave has zero effective angular momentum.  

\begin{figure}[h!]
  \centering

\end{figure}
\begin{figure}[h]
    \centering
    \begin{minipage}{0.5\textwidth}
          \includegraphics[width=0.8\textwidth]{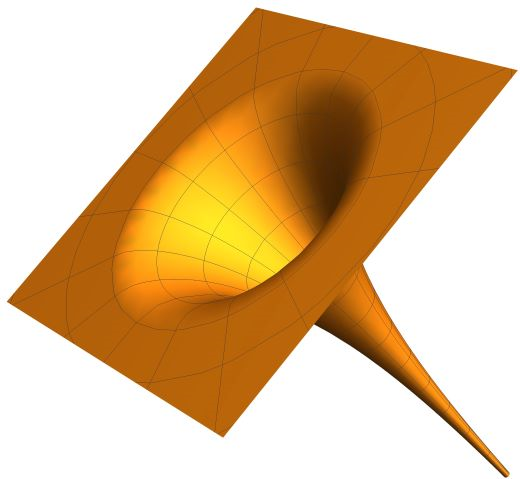}
    \end{minipage}\hfill
    \begin{minipage}{0.5\textwidth}
        \centering
        \includegraphics[width=0.8\textwidth]{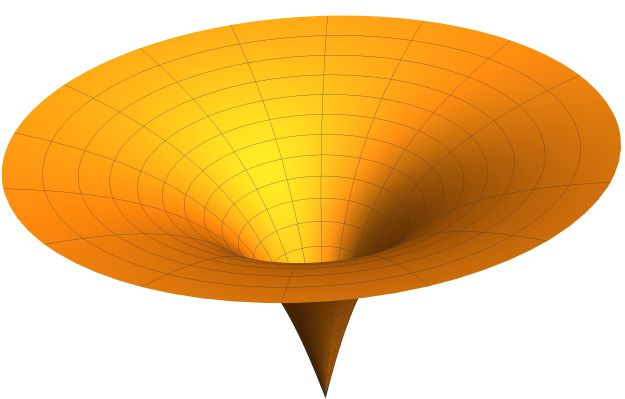}
    \end{minipage}
    \caption{Geometric structure of a Beltrami and elliptic wormhole surface. The Hilbert horizon(s) of Beltrami and elliptic wormhole is~(are) at $u_{H}=r\log\L[2r/a\R]$ and $u_{H}=r\log\Bigg( \displaystyle{\frac{r}{a}\pm\frac{r}{a}\sqrt{1- \frac{a^{2}}{r^{2}}}}\Bigg)$ respectively. }  \label{fig2}
\end{figure}

For the {\it constant field} scenario, the equation of motion (\ref{eomB0}) takes the form
\begin{equation}\label{Bel2}
\begin{split}
0
          &=\L[\partial_{u}^2
          +\frac{1}{r}\partial_{u}
          +\frac{1}{2r^{2}}-\L(\frac{1}{2r}\R)^2
          +2\frac{m\sigma}{ar}\e^{-u/r}-\L(2\frac{m}{a}\R)^2\e^{-2u/r}-\frac{a^2 e^{2 u/r}}{16 L^4}+\frac{a \sigma e^{u/r}}{4 L^2 r}+\frac{m}{L^2}+k^2\R] \varphi(u),
\end{split}
\end{equation}
where $L\equiv \sqrt{\hbar c/eB}$ is the magnetic length.  To solve (\ref{Bel2}), we approximate by considering the situation when the terms containing $a/L^{2}$ is negligible~(This is the limit where the magnetic length is larger than $a$. For typical $B=10 ~\text{T}, L\simeq 8.1 ~\text{nm}$, the wormhole throat radius $a$ needs to be smaller for the approximation to be valid).  The resulting equation of motion takes the form
\begin{equation}\label{Bel2a}
\begin{split}
0
          &=\L[\partial_{u}^2
          +\frac{1}{r}\partial_{u}
          +\frac{1}{2r^{2}}-\L(\frac{1}{2r}\R)^2
          +2\frac{m\sigma}{ar}\e^{-u/r}-\L(2\frac{m}{a}\R)^2\e^{-2u/r}+\frac{m}{L^2}+k^2\R] \varphi(u),
\end{split}
\end{equation}
which is exactly the same as (\ref{eomB0}) with replacement $m' \to m, k^{2}\to k^{2}+m/L^{2}$.  The solutions are thus the same with the replacement above.  Note that the energy formulae becomes
\be
\mp i \sqrt{k^{2}+m/L^{2}} ~r+\frac{1-\sigma}{2}=-n,
\ee
for non-negative integer $n$, leading to the energy quantization
\be
E^{2}_{n}-M^2c^4=k^{2}_{n} \hbar^2 c^2=-\frac{\hbar^2 c^2}{r^2}\Bigg\{ \left[ n+\Big( \frac{1-\sigma}{2}\Big) \right]^{2} +m \frac{r^{2}}{L^{2}}\Bigg\}.  
\ee
There is an orbit-magnetic coupling contributing to the energy.  The modes are normal for small $n$ and coupling but become QNMs with purely imaginary values for large coupling and/or highly excited states.

\subsubsection{Asymptotic solutions} \label{sec:AsymptoticBeltramiConFlux}

The similar analysis can be performed to explore the behaviour of solutions at both boundaries $u = 0, u_{H}$ in this $q=0$ case. We can express the equation of motion $(\ref{eq:Beltrami1})$ in the $Z$ coordinate as
\be
0 = Z^{2}\varphi''(Z)+\Big[ k^{2}r^{2}+\frac{1}{4}+\frac{\sigma}{2}Z-\Big(\frac{Z}{2}\Big)^{2} \Big]\varphi(Z),
\ee
that leads to
\begin{equation}\label{eq:AsymptoticBeltrami01}
\begin{split}
0=&\varphi''(Z)-\Delta^{2}_{0,H}\varphi(Z),
\end{split}
\end{equation}
in the asymptotic region $Z_{0,H}$ where $Z_{0}=4m'r/a$ and $Z_{H}=2m'$.

The solution takes the exponential form $\varphi(Z)\sim \e^{\Delta Z}$, with the asymptotic parameters given by
\begin{equation}\label{eq:AsymptoticBeltrami02}
\Delta_{0,H}=\pm \frac{i}{Z_{0,H}} \sqrt{\frac{1}{4}-\L(n+\frac{1-\sigma}{2}\R)^{2}+\frac{\sigma}{2}Z_{0,H}-\L(\frac{Z_{0,H}}{2}\R)^{2}},
\end{equation}
for the constant flux scenario.

\begin{table}[H]
\centering
\begin{tabular}{|c|c|c|}
 \hline
  &  $\Delta_{0}$ & $\Delta_{H}$ \\ 
 
 \hline
 
        &        &               \\ 
 
  const flux &  $\displaystyle{\pm \frac{i}{2m'} \frac{a}{2r} \sqrt{\frac{1}{4}-\L(n+\frac{1-\sigma}{2}\R)^{2}+2\sigma m'\frac{r}{a}-\L(2m'\frac{r}{a}\R)^{2}}}$  &  $\displaystyle{\pm \frac{i}{2m'}  \sqrt{\frac{1}{4}-\L(n+\frac{1-\sigma}{2}\R)^{2}+\sigma m'-m'^{2}}}$       \\

         &       &               \\ 

 \hline
 & \\
const field &  $\displaystyle{\pm \frac{i}{2m} \frac{a}{2r} \sqrt{\frac{1}{4}-\L(n+\frac{1-\sigma}{2}\R)^{2}+ m\L(2\sigma\frac{r}{a}-\frac{r^{2}}{L^{2}}\R)-\L(2m\frac{r}{a}\R)^{2}}}$  &  $\displaystyle{\pm \frac{i}{2m}  \sqrt{\frac{1}{4}-\L(n+\frac{1-\sigma}{2}\R)^{2}+m\L(\sigma-\frac{r^{2}}{L^{2}}\R)-m^{2}}}$       \\

         &       &               \\ 

 \hline
 \end{tabular}
\caption{The asymptotic parameter $\Delta_{0,H}$ at $u=0,u_{H}$ for Beltrami wormhole.}\label{Tabel:AsymptoticBeltramiConFlux}
\end{table}

For the constant field scenario, the asymptotic parameters are given by

\begin{equation}\label{eq:AsymptoticBeltrami04}
\Delta_{0,H}=\pm \frac{i}{Z_{0,H}} \sqrt{\frac{1}{4}-\L(n+\frac{1-\sigma}{2}\R)^{2}-m\frac{r^{2}}{L^{2}}+\frac{\sigma}{2}Z_{0,H}-\L(\frac{Z_{0,H}}{2}\R)^{2}}.
\end{equation}
They are summarized in Table~\ref{Tabel:AsymptoticBeltramiConFlux}.  Apparently, for low $n,m', m$~(but nonzero $m',m$) states, they are purely traveling waves in or out of the wormhole depending on the sign choices at both boundaries.

\subsection{Elliptic wormhole}

For elliptic wormhole with $q=-1$, all formulae of the hyperbolic cases can be used.  Notably since $\sqrt{q}=i$, all of the parameters $\alpha^{(\prime,\prime\prime)},\beta^{(\prime,\prime\prime)}$ become real and we can simply make replacement $\sqrt{q}\to i$ in the results of the hyperbolic cases, i.e. (\ref{E1}), (\ref{E2}) and (\ref{econB2}), (\ref{econB2n}).  The spin-orbit and all magnetic induced coupling terms become real.  

For constant flux scenario, states with low $n, m'$ usually are normal modes with real energy. This is consistent with the fact that the fermion cannot leak out through the throat~(since the throat size is zero in this case) and can only leak out from the Hibert horizon which requires large $n$ and/or $m'$. The QNMs only occur for highly excited states where the coupling terms are larger than the rest energy term.  When QNMs emerge, they are purely imaginary~(or diffusive) and contribute only in the form of dissipation.  Modes given by (\ref{E01}), (\ref{E02}) are not affected by the wormhole geometry, they are leaking solutions in the $u$ direction for highly excited states.  Angular-momentum-dependent modes (\ref{E1}) and (\ref{E2}) also give purely imaginary energy when the coupling terms are dominant.  Highly excited states with large $n$ could become normal modes if the coupling term $\sim \sigma m'$ has negative sign and cancel with the $(n+1/2)$ term under the square root.  This depends on the relative sign between the spin and $m'$.  Interestingly for fixed $n$, larger spin-orbit coupling $\sigma m'$ results in QNMs with shorter lifetime, i.e., larger $\text{Im}~E$.

For constant field scenario, modes given by (\ref{eq:54}), (\ref{eq:54n}) are independent of the wormhole geometry, they are normal modes for low $n,m$.  This is again consistent with the fact that the fermion cannot leak out through the throat.  It can only leak out from the Hibert horizon which requires large $n$ and/or $m$. For $m\geq 0$ they are leaking solutions~(purely dissipative modes) in the $u$ direction for sufficiently large $n$.  For $m<0$, the orbit-magnetic coupling could compensate with $(n+1)$ term resulting in the normal modes even for high $n,m$.  

Topologically, the elliptic wormhole is distinctively different from the hyperbolic and Beltrami ones.  The space starts at $u=0$ with $R=0$ so the modes cannot leak out through the hole, resulting in the absence of QNMs for low $n$ states in contrast to the hyperbolic and Beltrami cases.   

\section{Implications for graphene system}  \label{graSec}

The analyses and main results of our work are generic for any charged fermion spatially confined to the two-dimensional wormhole in the presence of the axial magnetic field.  The wormhole can be made from any kind of conductor, semimetal or semiconductor as long as the excited quasiparticles can move freely along the curved surface.  A special case worthwhile mentioning is the zero-gap semiconductor graphene where the electrons in the conducting band from the $2p_{z}$ orbitals behave like a massless fermion above the Fermi energy around the Dirac points in the momentum space.  Electrons in graphene can thus be described by a relativistc Dirac equation with replacement $c\to v_{F}\simeq 1\times 10^6$ m$/$s~(See e.g. \cite{Castro} for an excellent review).  In the continuum limit where the radius of the wormhole $a$ is much larger than the lattice size~\cite{Gonzalez}, we can make the following identification in the above analyses
\be
\varphi_{II}^{\sigma=+,-}=\varphi_{K}^{A,B},
\ee
where $K$ is one of the Dirac point $K, K'$ and $A,B$ are the two inequivalent atomic sites in the unit cell of the graphene lattice.  $\varphi_{K}^{A,B}$ are the corresponding wave functions of the fermion at each site.  The energy of quasiparticles around the Dirac point $K$ in the graphene wormhole can be calculated by setting $M=0, c=v_{F}$~({\it without} changing the definition of the magnetic flux quantum $\phi_{0}=hc/e$ and the magnetic length $L=\sqrt{\hbar c/eB}$.  In SI unit where the electric charge $e$ is measured in Coulombs, $\phi_{0}$ and $L$ do not originally contain $c$.) in the expression (\ref{E01}),(\ref{E02}),(\ref{E1}),(\ref{E2}) for constant flux as well as (\ref{eq:54}),(\ref{eq:54n}),(\ref{econB2}),(\ref{econB2n}) for constant field scenario.  The energy formulae can be summarized as the following

\underline{\bf Constant flux scenario~(exact)}
\begin{eqnarray}
E\left(\alpha=-\beta*=\frac{1}{4}+i\frac{\sigma m' r}{2a\sqrt{q}}\right) &=& \mp \left( \frac{\hbar v_{F}}{r}\right)\Big[\frac{\sigma r}{a\sqrt{q}}\left( m-\frac{\phi}{\phi_{0}} \right) - i \left( n+\frac{1}{2}\right)\Big], \label{Ereal1} \\
E\left(\alpha=-\beta*=-\frac{1}{4}-i\frac{\sigma m' r}{2a\sqrt{q}}\right) &=& \pm \left( \frac{\hbar v_{F}}{r}\right)\Big[\frac{\sigma r}{a\sqrt{q}}\left( m-\frac{\phi}{\phi_{0}} \right) + i \left( n+\frac{1}{2}\right)\Big], \label{Ereal2}\\
E\left(\alpha=\beta*=\frac{1}{4}+i\frac{\sigma m' r}{2a\sqrt{q}}\right) &=& \pm i  \left( \frac{\hbar v_{F}}{r}\right)\left( n+1\right), \label{Ediff1}\\
E\left(\alpha=\beta*=-\frac{1}{4}-i\frac{\sigma m' r}{2a\sqrt{q}}\right) &=& \pm i  \left( \frac{\hbar v_{F}}{r}\right)\left( n\right),  \label{Ediff2}
\end{eqnarray}
for integer $n$.   

\underline{\bf Constant field scenario~(approximate for $r<a$ or $X<1$)}
\begin{eqnarray}
E\left(\alpha'=-\beta'*=\frac{1}{4}+i\frac{\sigma m r}{2a\sqrt{q}}\right) &=&\pm i\frac{\hbar v_{F}}{r}\sqrt{\left(n+\frac{1}{2}+\frac{i\sigma m r}{a\sqrt{q}}\right)^{2}+mr^2 \frac{eB}{\hbar c} - \frac{iareB\sqrt{q}}{2\hbar c }}, \\
E\left(\alpha'=-\beta'*=-\frac{1}{4}-i\frac{\sigma m r}{2a\sqrt{q}}\right) &=&\pm i\frac{\hbar v_{F}}{r}\sqrt{\left(n+\frac{1}{2}-\frac{i\sigma m r}{a\sqrt{q}}\right)^{2}+mr^2 \frac{eB}{\hbar c} + \frac{iareB\sqrt{q}}{2\hbar c }}, \\
E\left(\alpha'=\beta'*=\frac{1}{4}+i\frac{\sigma m r}{2a\sqrt{q}}\right) &=&\pm i\frac{\hbar v_{F}}{r}\sqrt{(n+1)^{2}+(1+\sigma) mr^2 \frac{eB}{\hbar c}}, \label{Ediff3}\\
E\left(\alpha'=\beta'*=-\frac{1}{4}-i\frac{\sigma m r}{2a\sqrt{q}}\right) &=&\pm i\frac{\hbar v_{F}}{r}\sqrt{n^{2}+(1-\sigma) mr^2 \frac{eB}{\hbar c}}, \label{Ediff4}
\end{eqnarray}
for integer $n$.  The energies from (\ref{Ediff1}), (\ref{Ediff2}), (\ref{Ediff3}) and (\ref{Ediff4}) are purely imaginary implying that they are purely unstable~(exponentially growing) or decaying modes analogous to overdamped modes of oscillator.  The other modes have both real and imaginary parts of the energy and thus are quasinormal modes.

The energy scale of the fermion in the graphene wormhole is of the order of $\hbar v_{F}/r=0.658$ eV~nm$/r$ for $v_{F}=10^{6}$ m$/$s.  Naturally this is the same order as the electronic energy in the carbon nanotube with the similar radius and length.  Curvature effects, however, play a crucial role in the wormhole case where the imaginary energy is induced via the spin-orbit coupling and interaction between angular momentum and the external magnetic field.  Remarkably in the constant flux scenario at zero mass, the spin-orbit and spin-magnetic interaction energy become {\it real} as shown in (\ref{Ereal1}) and (\ref{Ereal2}).  In this case, the diffusive part of energy only comes from the leaking momentum out of the wormhole in the $u$ direction characterized by the quantum number $n$.  

The lifetime of the fermionic states of the graphene wormhole is characterized by $\tau=(\text{Im}~E)^{-1}$ which is also of the order of the inverse of $\hbar v_{F}/r=0.658$ eV~nm$/r$.  For $r=1-1000~\text{nm}$, $\tau =10^{-15}-10^{-12}$ secs, a considerably short period of time.  Interestingly for a macroscopic graphene wormhole of radius $100~\mu$m, the lifetime could be as long as $0.1$ nanoseconds.  For the constant flux scenario, the lifetime is purely determined by the quantity $\hbar v_{F}/r$ and independent of the angular momentum and the flux. For (\ref{Ediff3}) and (\ref{Ediff4}) of the constant field scenario, $m<0$ states in the presence of magnetic field can become stable with infinite lifetime when the quantity under the square root becomes negative and the energy takes the real values.  These are the Landau states.  Such long-lived states require sufficiently high magnetic field $B$ and low $n$ so that the Landau interaction is dominant. For e.g. $n=0, r=10$ nm, the magnetic field $B$ has to be larger than $6.58$ T so that $L>r$ for the dominant Landau interaction in (\ref{Ediff3}).

\section{Conclusions and Discussions}\label{sec:con}

We consider charged fermion in a two-dimensional wormhole in the presence of the external magnetic field with axial symmetry.  Assuming uniform field in the plane perpendicular to the direction of the field, we consider energy levels of fermion in two scenarios, constant flux through the wormhole throat and constant field.  The curvature connection of wormhole generates effective gauge connection resulting in the induced spin-orbit coupling of the fermion on the wormhole.  The coupling is genuinely ``gravitational'' since it exists even in the absence of the magnetic field and it is vanishing when the wormhole is flat, e.g. cylindrical wormhole.  When the magnetic field is turned on along the wormhole axis, the spin-orbit-magnetic coupling is also generated in addition to the conventional Landau coupling between the angular momentum of the fermion and the magnetic field.  This new interaction is the combined effect of gravity and gauge field on the charged fermion.  

A simple picture to help understanding these results is the following.  When a fermion is confined to the curved space like a two-dimensional wormhole considered here, its $\sigma^{3}$-spin component is perpendicular to the surface~(since the dreibein is locally defined in the tangent space of wormhole) while the orbital angular momentum is pointing along the $z$-direction.  The spin-orbit coupling $\sim \overrightarrow{\sigma^{3}}\cdot m\hat{z}$ is thus generated for generic wormhole with curvature.  For cylindrical wormhole tube, the surface is always parallel to $\hat{z}$ and the $\sigma^{3}$-spin component is always perpendicular to the surface so the spin-orbit coupling naturally vanishes.  

For both constant flux and constant field scenarios in every choice of solution parameters, sufficiently highly excited states with large $n$ will always give QNMs.  The energy naturally leaks out of the wormhole when the fermion is sufficiently excited.  This is consistent with the existence of Hilbert horizons~\cite{IoLa} at finite $u_{H}\equiv r\log\Bigg( \displaystyle{\frac{r}{a}\pm\frac{r}{a}\sqrt{1+ \frac{qa^{2}}{r^{2}}}}\Bigg)$ where the wormhole geometry ends.  Highly excited fermion lives at larger $u$ and it will leak out of the wormhole through the Hilbert horizons.  

On the other hand, the spin-orbit coupling always generate QNMs since the coupling~(on the wormhole) itself is imaginary $\sim i\sigma m$.  The origin of this term can be traced back to the pseudo gauge connection $A_{\tilde{u}}$ from Eq.~(\ref{coneq}) which is purely imaginary.  Remarkably, the curvature connection $\Gamma_{v}$~(i.e. ``gravity'') generates an effective (pseudo) gauge connection that is purely imaginary resulting in the complexity of the energy and the existence of the QNMs.  Physically, imaginary energy should be interpreted as the {\it energy dissipation} and {\it instability}.  Energy dissipation corresponds to the case with ${\rm Im}(E)<0$.  Instability stems from the enhancement in time of the wave function when ${\rm Im}(E)>0$.  A state with high orbital angular momentum $m$ tends to leak energy faster due to larger imaginary part of the QNMs. ์Note that the spin-orbit coupling term in the equation of motion is zero when $R'=0$~(at midpoint of the wormhole throat or in the case of cylindrical wormhole) and maximum when $R'=\cos\theta=1$ at the Hilbert horizon.  At Hilbert horizon, the surface is merging to the plane and perpendicular to $\hat{z}$.

The interplay between the curvature connection of the wormhole and the induced (pseudo) gauge connection demonstrates an interesting kind of gauge-gravity duality. The {\it real} gravity connection can be interpreted as the {\it imaginary}~(effective) gauge connection~(in the locally {\it perpendicular} direction on the surface) that leads to the complexity of the energy and the emergence of the QNMs and unstable modes.  Adding external magnetic field induces a new imaginary coupling term proportional to the field that only exists when there is curvature, i.e. last term in (\ref{Dsq}) and (\ref{Dsq1}).  The new curvature-spin-magnetic field coupling similarly leads to the emergence of QNMs and unstable modes.

The energy and lifetime of charged fermion in the graphene wormhole are determined by the characteristic energy $\hbar v_{F}/r=0.658$~eV~nm$/r$.  For typical wormhole with radius $r=100~\mu\text{m}$, the lifetime $\tau=0.1$ nanoseconds.  For constant field scenario with choices of states given by (\ref{Ediff3}) and (\ref{Ediff4}), states with angular momentum in the opposite direction to the magnetic field can become stable with infinite lifetime.

The gauge field in the wormhole can change the total angular momentum of the charged fermions, altering their statistics accordingly~(see Ref.~\cite{Wilczek} for discussion in cylinder). The effective orbital quantum number is given by $m'=m-\phi/\phi_{0}$.  Since the magnetic flux quantum is $\phi_{0}=hc/e=4.13567\times 10^{-15}$ Tm$^2$: for $B=10$ T and $a=\sqrt{\phi_{0}/2\pi B}=8.11$ nm, we then have $m'=m-1/2$ and the total angular momentum becomes integer.  The fermion quasiparticle~(e.g. electron or hole) would behave like a boson {\it with charge $e$, the electron charge}. It should be emphasized that these boson-like fermions are not pairs of fermions like the Cooper pairs, their statistics are simply altered by global boundary condition in the presence of the magnetic field. It is possible to store condensated boson-like~(when the flux is half-integer of $\phi_{0}$) fermions in the wormhole connecting e.g. two graphene sheets and control their behaviour by changing either the magnetic field or the shape of the wormhole. This could potentially lead to a number of profound electronic properties and future applications.   

On the other hand, if the constant flux is trapped within a vortex such as in the type II superconductors, the charged particle in this case is the Cooper pairs with charge $2e$ which is a boson satisfying the Klein-Gordon equation given also by Eq.~(\ref{eq:25}).  The Cooper pairs on the vortex surface in this case will remain boson for integer values of $2\phi/\phi_{0}$~(Cooper pairs have $2e$ charge). In the interior of the vortex where the flux is smaller, the statistics become arbitrary~(non-Bose-Einstein statistics) and the Cooper pairs cannot condensate.  In this way, we can conclude that there is a minimum magnetic flux~(above zero) when $2\phi/\phi_{0}=1$, i.e., the conventional magnetic flux quantum $\phi=\phi_{0}/2=hc/2e$, below which the Bose-Einstein condensation cannot occur and the region within this vortex is in the non-superconducting phase~(superconducting phase can occur for zero flux as long as the angular momentum of the charged particles is also zero, i.e., no vortex). This is one way to argue the existence of the magnetic flux quantum in type II superconductor.

Generically, we can calculate the radius $a$ at which statistics of quasiparticles exchange between fermion and boson to be
\be
a = \sqrt{\frac{\hbar c}{eB}}\sqrt{2n+1}=L\sqrt{2n+1}, \quad n=0,1,2,... .  \label{radeq}
\ee
For graphene wormholes at these radii in the constant flux scenario, boson-like fermions should condensate at the ground state within the hole at sufficiently low temperature $T<\hbar v_{F}/k_{B}a=7639$ K nm/a.  To achieve condensation at room temperature $T=300$ K, we need $a\lesssim 25.5$ nm implying also from Eq.~(\ref{radeq}) that $B\gtrsim 1$ T for $n=0$.

\section*{Acknowledgments}  T.R. would like to thank Sutee Boonchui for support while he was studying at the Kasetsart University as well as previous collaboration on the related topic.  P.B. is supported in part by the Thailand Research Fund~(TRF), Office of Higher Education Commission (OHEC) and Chulalongkorn University under grant RSA6180002. 


\end{document}